\documentclass{aa}

\newcommand{\ttend}{\tau_\tend}

\newcommand{\tanti}{{\text{anti}}}
\newcommand{\tsym}{{\text{sym}}}

\newcommand{\tdecay}{{\text{decay}}}
\newcommand{\tviscous}{{\text{visc}}}
\newcommand{\tAlfven}{{\text{A}}}
\newcommand{\tOhm}{{\text{Ohm}}}

\newcommand{\ttdecay}{\tau_\tdecay}
\newcommand{\tgrow}{\tau_{\prec}}
\newcommand{\tdec}{\tau_{\succ}}
\newcommand{\ttA}{\tau_\tAlfven}
\newcommand{\ttAO}{\tau_{\tAlfven,0}}
\newcommand{\ttOhm}{\tau_\tOhm}

\newcommand{\tend}{{\text{end}}}
\newcommand{\Alf}{Alfv\`en}

\usepackage{graphicx}
\usepackage{epic}
\usepackage{array}
\usepackage{curves}
\usepackage{amscd}
\usepackage{amsmath}
\usepackage{amssymb}
\usepackage{multirow}
\usepackage{natbib}
\usepackage{float}

\newcommand{\deriv}[3]{\frac{#3\hspace*{-.06em} {#1}}{#3\hspace*{.06em} {#2}}}

\newcommand{\parder}[2]{\deriv{#1}{#2}{\partial}}

\newcommand{\rezip}[1]{\frac{1}{#1}}

\newcommand{\curl}{\operatorname{curl}}
\newcommand{\dv}{\operatorname{div}}

\renewcommand{\exp}{\operatorname{exp}}

\newcommand{\undertilde}[1]{\mbox{$\underset{\displaystyle\hspace*{-.06em}\widetilde{}}{#1}$}}
\newcommand{\ovec}[1]{{\mbox{\boldmath $#1$}}}
\newcommand{\rvec}{\ovec{r}}

\newcommand{\ervec}{\ovec{e}_r}

\newcommand{\ezvec}{\ovec{e}_z}
\newcommand{\zervec}{\ovec{0}}
\newcommand{\Ynm}{Y_n^m}

\newcommand{\tA}{{\text{A}}}

\newcommand{\sumeu}[1]{\sum_{#1=1}^\infty}

\newcommand{\Bvec}{\ovec{B}}

\newcommand{\uvec}{\ovec{u}}
\newcommand{\Uvec}{\ovec{U}}

\newcommand{\Pm}{\operatorname{\mathit{Pm}}}

\newcommand{\tmag}{{\text{mag}}}
\newcommand{\tkin}{{\text{kin}}}
\newcommand{\Emag}{E_\tmag}
\newcommand{\Ekin}{E_\tkin}

\newcommand{\Pimag}{\Pi_\tmag}
\newcommand{\Pikin}{\Pi_\tkin}
\newcommand{\Mmag}{M_\tmag}
\newcommand{\Mkin}{M_\tkin}

\newcommand{\tP}{{\text{P}}}
\newcommand{\tT}{{\text{T}}}

\newcommand{\ntoro}[1]{\rvec\!\times\!\nabla #1}

\newcommand{\laBi}{\lambda_i^B}
\newcommand{\laUi}{\lambda_i^U}

\newcommand{\myref}[1]{~\hspace{0pt plus 1pt minus 1pt}\ref{#1}}

\newcommand{\sectref}[1]{Sect.\myref{#1}}

\newcommand{\figsandref}[2]{Figs.\myref{#1} and\myref{#2}}
\newcommand{\figsref}[2]{\figsandref{#1}{#2}}

\newcommand{\figref}[1]{Fig.\myref{#1}}
\newcommand{\tablref}[1]{Table\myref{#1}}
\newcounter{saveqn}
\newcommand{\alpheqn}{\refstepcounter{equation}\setcounter{saveqn}{\value{equati
on}}%
\setcounter{equation}{0}%
\renewcommand{\theequation}{\mbox{\arabic{chapter}.\arabic{saveqn}\alph{equation
}}}}
 
\newcommand{\reseteqn}{\setcounter{equation}{\value{saveqn}}%
\renewcommand{\theequation}{\arabic{chapter}.\arabic{equation}}}

\begin{document}
 \title{Magnetars versus Radio Pulsars:}
 \subtitle{MHD Stability in Newborn Highly Magnetized Neutron Stars}
 \titlerunning{Magnetars versus Radio Pulsars}
  \author{ U. Geppert\inst{1} \and M. Rheinhardt\inst{2} }
  \institute{Max--Planck--Institut f\"ur Extraterrestrische Physik, Gie{\ss}enbachstra{\ss}e,
     85748 Garching, Germany, \email{urme@xray.mpe.mpg.de}
     \and
     Astrophysikalisches Institut Potsdam,
      An der Sternwarte  16, 14482 Potsdam, Germany, \email{MReinhardt@aip.de}
                 }
\offprints{U. Geppert}

\date{Received ; accepted }

\abstract{}%
{We study the stability/establishment of dipolar magnetostatic equilibrium configurations in new--born
neutron stars (NSs) in dependence on the rotational velocity $\Omega$ and on the initial angle
$\alpha$ between rotation and magnetic axis.}%
{The NS is modeled as a sphere of a highly magnetized ($B \sim 10^{15}$G) incompressible fluid of uniform density which rotates rigidly.  For the initial dipolar background magnetic field, which defines the magnetic axis, two different configurations are assumed.
We solve the 3D non--linear MHD equations by use of a spectral code. The problem in dimensionless form is completely defined by the initial field strength (for a fixed field geometry), the magnetic Prandtl number $\Pm$, and the normalized rotation rate. The evolution of the magnetic and velocity fields is  considered for
initial magnetic field strengths characterized by the ratio of ohmic diffusion
and initial \Alf{} travel times  $\ttOhm/\ttAO \approx 1000$, for $\Pm = 0.1, 1,
10$, and  the ratio of rotation period and initial \Alf{} travel time, $P/\ttAO = 0.012, 0.12, 1.2, 12$.}%
{We find hints for the existence of a unique stable dipolar magnetostatic configuration for any specific $\alpha$, independent of the initial field geometry. Comparing NSs possessing the same field structure at the end of their proto--NS phase, it turns out that sufficiently fast rotating NSs ($P\la6\,$ms) with $\alpha \la 45^0$ retain their magnetar field, while the others lose almost all of their initial magnetic energy by transferring it into magnetic and kinetic energy of relatively small--scaled fields and continue their life as radio pulsars with a dipolar surface field of $10^{12\dots13}$G.}{}

\keywords{Magnetohydrodynamics; Instabilities; Stars: magnetic fields; Stars: neutron}

\maketitle

\section{Introduction}

The strength of a neutron star's (NS's) dipolar  magnetic field, its inclination angle with respect to the
NS's rotation axis, $\alpha$, and the rotational velocity $\Omega$ are the 
quantities determining the
observability of pulsars as well as the power of their magneto--dipole radiation and/or the pulsar wind, which
eventually set the spin--down behavior of an isolated pulsar \citep{ST83}.
Therefore, the determination of
these (more or less exactly observable) quantities in their temporal evolution is subject of scientific debates since the discovery of pulsars.

From the observational point of view there exists only little information about the magnetic inclination $\alpha$ immediately after birth of a NS.
This is owed to the poor observability of new--born NSs, which are surrounded by opaque material.
Recently, \citet{HAGC03} and \citet{H06} investigated the emission geometry of very young and energetic
pulsars and found that five out of six have $\alpha \lesssim 30^\circ$.
The 1600 years old pulsar \object{J1119-6127}, e.g., has an inclination angle $\alpha = 19^\circ$ and a dipolar magnetic field strength of $4.2\cdot 10^{13}$G.
Age and dipolar magnetic field strength correspond to the observed spin period of $0.42$ s, provided the field has been constant during the lifetime of the NS.
Given that constancy an initial spin period of less than a millisecond can be estimated \citep[see, e.g.,][]{CGP00}.

Unfortunately, there is even less observational evidence about $\alpha$ in magnetars
because their slow rotation impedes the detection of open field line regions.
The typically wide X--ray pulses come likely from a very wide beam.
If the X--ray emission region is not too far above the surface and the pulse peak corresponds to the magnetic pole, then that wide beam may indicate that the inclination angle $\alpha$ is small \citep{Z06}. 

In order to understand those, yet vague, observational evidences of relatively small values of $\alpha$ for magnetars and young energetic pulsars, and to get at all an idea about the ``initial'' magnetic field structure of NSs, the study of the establishment/stability of magnetostatic equilibrium configurations immediately after a NS's birth is necessary.
In this very early phase, when the core is not yet superfluid,
consists of npe--matter, and when the crust is completely liquid, an equilibrium configuration may indeed be reached.
It defines the initial magnetic field structure and strength, especially that of the dipolar component, which plays a decisive role for spin--down behavior and observability of the NS.

In new--born NSs, the rotation velocity exceeds the hydromagnetic (or \Alf{}) one, $v_\tAlfven$, in large regions, even for magnetic field strengths in the order of magnetar fields ($B \lesssim 10^{15}$G).
Therefore, the effect which rotation may have on the establishment of a magnetostatic equilibrium in young NSs has to be studied in detail.

It was \citet{W73}, who postulated the stability of a magnetic field configuration the poloidal and toroidal constituents of which are of comparable strength.
This conclusion was supported by subsequent studies, e.g., \citet{MT73} or \citet{BN05}.
Tayler and co--authors performed a series of analytical stability analyses of magnetic fields in stars:
\citet{T73} considered the stability of a purely toroidal axisymmetric field without rotation and found that it is always unstable with respect to azimuthally large--scale ($m=1$) perturbations.
The growth rate of this instability  is on the order of the \Alf{} frequency $\Omega_A= v_A/r$, where $v_A = B/\sqrt{4\pi\rho}$, $B$ the magnetic field strength,  $\rho$ the density and $r$ the radial coordinate.
This ``Tayler--instability'', meanwhile basic ingredient of the ``Spruit--dynamo'' \citep[]{S99, S02}, acts locally in meridional ($r,\theta$) planes but globally in the azimuthal ($\varphi$) direction.
\citet{MT73} and \citet{W73} performed similar adiabatic stability analyses for an axisymmetric poloidal field and concluded its general instability if at least some of its field lines are closed within the star. The instability generates small scale field structures on the same time scale as for a toroidal field which dissipate rapidly. \cite{MT73}  discuss the possibility that sufficiently rapid rotation and/or the simultaneous existence of equipollent  toroidal and poloidal components may stabilize the axisymmetric background fields. 

\citet{PT85} explored in ideal MHD and cylindrical geometry the effect of rigid rotation on the stability of a field consisting of a toroidal component which increases with the distance from the magnetic axis, and/or a uniform $z$ (i.e., poloidal) component for different analytically feasible special cases: 
For $\alpha=0$ both the purely toroidal field  and the combined toroidal--poloidal one show qualitatively the same stabilizing effect of the rotation.
For $\alpha=90^\circ$, however, a purely toroidal field reveals a much less stabilizing effect of the rotation.
They studied further the stabilizing effects of a stable stratification of the star's density profile in the case $\alpha = 0^\circ$ and for a purely toroidal field which may increase non--linearly with axial distance.
As already shown by \citet{T57}, such a field is stable even in the absence of rotation for sufficient small wavelengths of a perturbation along the magnetic axis. Larger scaled perturbations, however, cause instabilities which can be counteracted by sufficiently rapid rotation.
On the other hand, \citet{PT85} found that at the same rotation rate, which stabilizes the original instabilities, rotation introduces a new instability. Therefore, they concluded, that ``rotation is unlikely to lead to complete stability of general magnetic field configurations''.

In 2005 the interest in the stability of magnetostatic equilibria in stars became revitalized by a series of papers: \citet{BS05a,B05,BN05}.
The latter authors  studied the stability of magnetic fields in the radiative interiors of Ap
stars and found, that  any random field is generally unstable but evolves in a stably stratified star towards a ``twisted torus'' configuration with approximately equipollent toroidal and poloidal components.
\citet{BS05a} considered the stability of MHD configurations in ultra--magnetized NSs, the magnetars.
Although they found that, quite similar to the Ap stars, stable magnetic fields
exist when being concentrated to a relatively small region around the center of
the star and after having evolved into a poloidal--toroidal torus shape, these fields cannot have magnetar strength unless their initial strength exceeds 10$^{18}$ G.
This estimate relies on two field--diminishing effects in the simulations of \citet{BN05}:
Firstly, a decrease of typically $\lesssim 10$ appears during the relaxation into the torus--like structure; 
secondly, these fields are strongly concentrated in the central region and decay by a factor of up to 1000 towards the surface.
However, NSs born with magnetic fields of $\sim10^{18}$ G are unlikely.
On the other hand an internal field of (only) magnetar strength would need a much too long time to diffuse to the surface. 

Until now the effect of rotation on the stability of purely poloidal field configurations has not been considered in general.
The results of \citet{PT85}, \citet{L88} and \citet{CP95} do not apply directly 
since these works consider combined toroidal and poloidal fields, the latter being homogeneous.
In the sole case in which \citet{CP95} consider an inhomogeneous poloidal field, the Tayler instability
was not found. 
The work of Braithwaite and Spruit doesn't include rotation except in \citet{B05} who considers toroidal fields only.
 
Since rotation in new--born NSs can for sure not be neglected and the dipolar magnetic field determines both observability and rotational evolution, we will study the consequences of rotation for the stability of such fields.
In the next section we describe and justify our model assumptions, present the set of equations to be solved together with the initial conditions for the models under consideration and sketch our numerical approach.
After presenting the results for the different MHD configurations in \sectref{results}, we finally discuss the results.

\section{Model, basic equations, numerical method}
\subsection{Description of the model}
We study the stability of magnetostatic configurations in a model applicable to strongly magnetized, rotating new--born NSs which have just established their (almost) stationary density profile according to the thermodynamic properties of NS matter (expressed by an equation of state).
It is a highly conductive normal--fluid npe liquid considered incompressible, a perhaps not too crude approximation for young NSs.
Furthermore, we assume a constant density profile and rigid rotation throughout the star, that is, all internal motions to be damped out initially.

Let us discuss these model assumptions in more detail:
\begin{enumerate}
\item
We assume that the crust is still completely liquid and that the core matter has not yet performed the phase transition into the superfluid state.
The onset of superfluidity and/or crystallization limits the period during which a
magnetostatic equilibrium can establish undisturbedly to at most a few hours.
Although the temperature of the transition into the superfluid phase, $T_\text{c}$, is rather
uncertain, it probably lies in wide regions of the NS between $10^9$ and $10^{10}$K.
According to the approximate analytical formulae given in Sect. 3 of \cite{PGW05},
the period after which the NS has cooled down to these temperatures is $ 17$ s ... $0.5$ yrs.
On the other hand, a characteristic \Alf{} time is on the order of $\ttA \approx 0.02\ldots 20$ s, depending on the background field strength.
Wisely assuming that $T_\text{c} \la 8\cdot 10^9$K for the majority of non--exotic equations of state, we believe that there is ample time for the MHD configuration to reach either a stable state or to go through an unstable phase dissipating magnetic energy very efficiently \citep[see][]{BS05a}.
The same is valid for the onset of crystallization at the crust--core interface.
A typical value for the melting temperature at the density there, $\rho \approx 1.5 \cdot 10^{14}$g cm$^{-3}$, is $T_\text{m} \approx  8\cdot 10^9$ K \citep {J99} so that about the same cooling period may elapse until the onset of crystallization in the crust as passes until the transition into the superfluid state in the core and inner crust.
Hence, using again the above mentioned formulae, for at least $1000$ seconds after the NS's birth the state of its matter is that of a normal--fluid npe liquid.
\item
At the end of the PNS phase a very strong magnetic field is present in the star, either due to PNS dynamo action \citep[see, e.g.,][and references therein]{RG05} or simply by flux conservation during the collapse.
The field may comprise components of very different  scale lengths, but is likely to possess a strong, large scale, say, dipolar fraction which defines the magnetic axis of the star.
\item
It is very likely, too, that the NS is rapidly rotating immediately after its birth. Since braking effects had not yet time to act efficiently, it follows from angular momentum conservation \citep[see, e.g., model calculations of the progenitor's rotation in ][] {HWS05} that the initial spin period of the NS lies in the range of $1\ldots10$ ms.
\citet{OBTLW05} studied the evolution of rotational profiles in proto--NSs taking into account magnetic effects in parameterized form.
They found that a proto--NS rotating differentially at the beginning will  reach very fast the state of rigid rotation by angular momentum redistribution due to turbulent (magnetically induced) viscous dissipation, so that at the end of the proto--NS phase, when the definite, non--convective, configuration of the new--born NS is reached, it rotates most likely rigidly \citep[see also][]{S00}. 
\item
In order to let the assumption of incompressibility be justified the fluid velocity should be well below the sound velocity.
We estimate the latter to be in the range
$c_s =8 \cdot 10^8 $cm s$^{-1}\ldots \lesssim$ speed of light  for the density ranging from
$10^{12}$ to $3\cdot 10^{16}$g cm$^{-3}$ according to \citet{BPS71}.
Of course, the results of our model
have to be checked {\it a posteriori} to be consistent with this requirement.
The same has to be done with respect to the criterion that the propagation time of sound waves across the star must be short in comparison with the time scales of 
the velocity.
\item
The assumption of constant density is surely less well justified.
However, for non--exotic equations of state \citep[see][]{BPS71} the change in density is only about one order of magnitude within 
three quarters of the NS volume.
Again {\it a--posteriori}, it has to be checked whether the bulk of the flow is concentrated within that core volume. 
As a future refinement of the model, realistic density profiles should be included in the framework of the anelastic approximation.
\item
As both the magnetic and kinetic energies are small in comparison with the gravitational binding, any deviation of the NS from
sphericity is neglected.
Another  approximation consists in the assumption of a perfect vacuum in the NS's surroundings which implies necessarily a
vanishing radial component of the velocity at the surface.
Both approximations together imply that only two of the three components of the force on a surface element can be required to be zero (see Eq. \eqref{sfRB}).
\end{enumerate}

\subsection{Basic equations}
We decided to solve the non--linear MHD equations as presented below although a linear
analysis would be sufficient to detect instabilities. Hence, we solve an
initial--value problem where the background field together with some arbitrary
perturbations form the initial condition.
Since we imprint only perturbations whose energy is very small in comparison
with that of the background field, there is always a linear stage from which 
the same growth/decay rates  can be derived as from a stability analysis based
on the linearized equations. The advantage of solving the non--linear equations
consists in the additional informations we obtain with respect to the duration
of the linear stage and the saturation of the ``final'' MHD configuration which
may represent another, now stable, equilibrium. As the symmetry of  the fastest
growing eigenmodes of the perturbations is {\it a--priori} unknown \citep[cf., e.g.,][]{MT73} we have to solve in 3D without any symmetry constraints.

The preceding considerations result in the following ruling MHD equations for magnetic field $\Bvec$ and velocity $\uvec$ and corresponding boundary conditions  written in dimensionless form:
\begin{equation}
           \begin{alignedat}{2}
               \parder{\Bvec}{t} &= \Delta\,\Bvec + \curl(\uvec\times\Bvec) &\qquad\text{for} \quad r \le 1 \\
                \curl \Bvec &= \zervec  &\qquad\text{for} \quad r > 1 \\
               \dv\Bvec &= 0   &\qquad\forall r                                                      \\
               [\Bvec] &= \zervec &\qquad\text{for}\quad r=1
            \end{alignedat}\label{ind}
\end{equation}
\begin{align}
        &\hspace*{-4mm} \left.\begin{alignedat}{3}
                \hspace*{-1.5cm}\parder{\uvec}{t} &=  &-&\,\nabla( p - \rezip{2}\Omega^2 r^2 \sin^2\vartheta+\Phi ) \\
                                                                                  &  &- &\,(\uvec\nabla)\uvec  + \Pm\Delta\,\uvec \\
	                                                                          &  &- &\,2 \Omega \ezvec\times\uvec + \curl\Bvec\times\Bvec\\ 
                \dv\uvec&= \;0   
           \end{alignedat}\,\right\}\;\; \text{for}\quad r \le 1\hspace{-3mm} \label{mom}\\*[0.5mm]
         &\hspace*{1cm}  \left.\begin{aligned}
                 u_r &= 0 \\
	        (\undertilde{D}(\uvec)\cdot\ervec)_{\vartheta,\varphi}& = 0 
            \end{aligned}\;\right\}  \qquad\text{for} \quad r=1\hspace*{-3cm}\label{sfRB}
\end{align}
 Here, length is normalized on the NS radius $R$, time on the magnetic diffusion time
 $\ttOhm=R^2/\eta$, where $\eta=c^2/4\pi\sigma$  is the magnetic diffusivity and $\sigma$ the electric
 conductivity. $[.]$ denotes the jump of a quantity across a surface.
The magnetic field is measured in units of $ (4\pi\rho)^{1/2}\eta/R$, the velocity in units of $\eta/R$,
the pressure $p$ in units of $\rho\eta^2/R^2$, and the gravitational potential $\Phi$ in units of $\eta^2/R^2$.
$\undertilde{D}(\uvec)$ denotes the deformation tensor.
$(r,\vartheta, \varphi)$ are spherical co--ordinates, the polar axis ($\vartheta=0$, $\upuparrows \ezvec$) of which coincides with the axis of rotation.
 
Equations \eqref{ind} and \eqref{mom} contain two parameters which together with the initial field strength $B_0$ (for a fixed field geometry) define the problem 
completely: the magnetic Prandtl number $\Pm$ and the normalized rotation rate $\Omega$.

$\Pm$, is the ratio of kinematic viscosity, $\nu$, and magnetic diffusivity, $\eta$,
and is chosen to be 0.1, 1, and 10, resp., what is partly dictated by numerical restrictions,
but on the other hand represents a subset of the NS--relevant range determined by the prevailing densities ($\rho= (1\ldots2.8)\cdot10^{14}$ gcm$^{-3}$) and temperatures ($\sim 10^{11\ldots9} $K).
These values result in $\Pm =0.1 \ldots 6\cdot10^8$ \citep{CL87,YS91} reflecting mainly the strong 
temperature  dependence of $\Pm\propto T^{-4}$. 

As we rely on a dimensionless representation, the two remaining parameters are best expressed as ratios of  characteristic times:
$\Omega$ by $q_P=P/\ttAO$, $P$ the rotation period of the NS, and the initial field amplitude $B_0$ by $q_{B_0}=\ttAO/\min\{\ttdecay^B,\ttdecay^u\}$, where $\ttAO= R/v_{\tA,0}$ is the \Alf{} crossing  time related to the initial field.
Note, that the ``decay times'', $\ttdecay^B$, $\ttdecay^u$, are different from the
often used magnetic and viscous ``diffusion times'', $\ttOhm=R^2/\eta, \; \tau_\tviscous =  R^2/\nu$, by factors $1/\pi^2$ and $0.06676$, respectively.
In the following we use $\ttdecay$ for $\min\{\ttdecay^B,\ttdecay^u\}$. 

For $q_{B_0}$ we take values $\sim 10^{-2}$, which are of course by far too large for highly magnetized young NSs where this ratio can go down to $10^{-16}$ for a surface magnetic field of $10^{15}$ G. However, as long as the growth/decay times of the examined perturbations of the initial state are in the order of at most a few \Alf{} times, dissipation of the background state does not  affect their linear stage (quasi--stationary approximation). 

The artificially enhanced dissipation is a concession to numerical feasibility only, in order to avoid excessive requirements for spatial resolution
\citep[cf. ][ where $q_{B_0}$ was chosen to be 0.1]{BS04}. Hence, the use of the results from the nonlinear stage has to be considered with caution.
(see \sectref{disc}).

For all models in the present study we choose the following parameters
\begin{equation}
R=10^6\text{cm}\, ,\:
\rho=2\cdot10^{14}\text{g/cm}^3\, ,\:
\tilde{B}(t=0)=10^{15} \text{G},
\label{denorm}
\end{equation}
where $\tilde{B}$ is the de--normalized r.m.s. value. Therefore, $\ttAO=0.05$ s. With  rotation periods  chosen between 0.6 ms and 0.6 s,  where the former value is close to the generally accepted minimum  for new--born NSs, but the latter is in that respect not very likely, these model parameters result in $q_P=0.012\ldots12$. As a reference and for comparison with known results for non--rotating stars we also consider $q_P=\infty$.

\subsection{Initial conditions}
Each initial condition consists of an axisymmetric dipolar background field the stability of which is examined, and
imprinted magnetic perturbations. The fluid is assumed to be at rest
initially.
The initial magnetic inclination $\alpha$ is defined by the dipole axis.
Of course, for $\alpha\ne0$ the background fields shown in \figsref{fig:uniform}{fig:poleq} have to be considered
properly rotated. 
The energy of the perturbations is set to 0.1\% of the magnetic background energy, and their chosen
geometry ensures that the initial state has no preferred equatorial symmetry. 
Their azimuthal dependence is characterized by $m=6,7$. This apparently arbitrary choice is motivated by numerical feasibility but at least doesn't affect the validity of the results  for the linear stage.
\subsubsection{Internal uniform field}\label{init_florud}
 \begin{figure}[h]
    
    \vspace{-.7cm}
     \hspace*{0cm}\includegraphics[width=.5\textwidth]{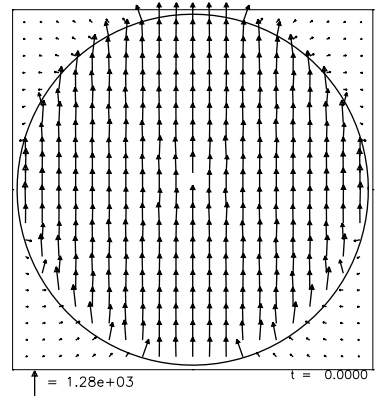} 
    \caption{Internal uniform field according to \eqref{uniform} in a meridional plane. The arrow indicated below marks a magnetic field of $10^{15}$G. }
    \label{fig:uniform}
 \end{figure}
As a first, more academic example we consider a field uniform throughout the sphere, but being a dipolar vacuum field outside: 
 \begin{equation}
\begin{alignedat}{2}
\Bvec &= \phantom{-}B_0\ezvec &\quad \text{for}\; r\le 1\\
\Bvec &= - B_0 \curl\rvec\times\nabla\left (\frac{\cos\vartheta}{2r^2}\right)&\quad \text{for}\; r > 1\\
[B_r] &= 0  &\quad \text{for}\; r= 1
 \end{alignedat}\label{uniform}
 \end{equation}
 Here, $B_0$ denotes the polar surface field strength. Of course, the continuity of the normal component of $\Bvec$ has to be required, but the tangential components remain discontinuous and give rise to surface currents.
 We have to concede, however, that the used numerical method enforces continuity of all components of $\Bvec$ on the 
 surface whereby we in fact used a field which in a thin shell immediately below the surface deviates from homogeneity (see \figref{fig:uniform}).
We include this model field into our study because it was the one
considerations on magnetic stability in NSs were first exemplified on by
\citet{FR77} and because \citet{BS05a} reported on its instability in the non--rotating case.

\subsubsection{Dipolar magnetostatic equilibrium field}
 \begin{figure}[htbp]
    
    \vspace{-.5cm}
     \hspace*{0cm}\includegraphics[width=.5\textwidth]{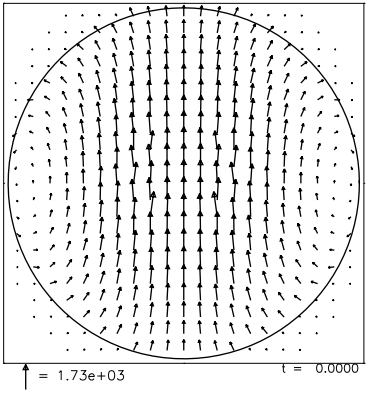} 
    \caption{Dipolar magnetostatic equilibrium field according to \eqref{poleq}. The arrow indicated below marks a magnetic field of $1.71\cdot10^{15}$G.}
    \label{fig:poleq}
 \end{figure}
A more realistic initial configuration consists in the poloidal magnetostatic equilibrium field 
the angular dependence of which is the same as for a dipolar field \citep[][see \figref{fig:poleq}]{R81,RKG04}:   
 \begin{equation}
\begin{alignedat}{2}
\Bvec &= \phantom{-}B_0\;\curl\rvec\times\nabla\left (\frac{1}{4}(3r^3-5r) \cos\vartheta\right)&\quad \text{for}\; r\le 1\\
\Bvec &= - B_0\; \curl\rvec\times\nabla\left (\frac{ \cos\vartheta}{2r^2}\right)&\quad \text{for}\; r > 1\\
[\Bvec] &= 0 & \quad \text{for}\; r= 1
 \end{alignedat}\label{poleq}
 \end{equation}
 As the Lorentz force of this field is a pure gradient it can in the constant--density case be balanced by the quantity $-\nabla\big( p - (\Omega^2 r^2 \sin^2 \vartheta)/2 + \Phi \big)$ (see Eq. \eqref{mom}). If stable, this state could be a final equilibrium to which an arbitrary initial configuration relaxes during the fluid stage of a newly born NS. As known from analytical studies, however, \citep{MT73} any purely poloidal field is in the absence 
 of rotation surely unstable, hence the evolution of the field \eqref{poleq} under the influence of rotation is a crucial question.
\subsection{Numerical method} \label{method}
Because the star is modeled as a  spherical body and incompressibility is assumed, the problem is especially suited to be tackled 
by a spectral method. The one used here employs an expansion of $\Bvec$ and $\uvec$ into their modes of free decay $\Bvec_i(\rvec)$,
$\uvec_i(\rvec)$. The $\Bvec_i(\rvec)$ are
the time--independent parts of the eigensolutions of  \eqref{ind} with $\uvec=\zervec$. Similarly, the $\uvec_i(\rvec)$ are obtained as the time--independent parts of the eigensolutions of \eqref{mom}, \eqref{sfRB}  with $\Bvec=\zervec$, $\Omega=0$, and the term $(\uvec\nabla)\uvec$ canceled. It can be shown, that both sets of modes form orthogonal systems of
vector functions, which are complete with respect to all vector functions obeying the respective boundary conditions and 
being solenoidal \citep[see, e.g.,][]{R98}. That is, the solutions can be represented as the infinite sums 
\begin{equation}
        \begin{aligned} 
              \Bvec(\rvec,t) &= \sumeu{i} b_i(t) \Bvec_i(\rvec)  \;,\\
              \uvec(\rvec,t) &= \sumeu{i} u_i(t) \uvec_i(\rvec)
            \end{aligned}\label{expan}
\end{equation}
with	    
\begin{equation}
            \begin{alignedat}{3}
              \Delta\,\Bvec_i + \laBi\Bvec_i &= \zervec\,,\; &\dv \Bvec_i&=0\\*[1mm]
              \Delta\,\Uvec_i + \laUi\Uvec_i &= \nabla p_i\,,\; &\dv \uvec_i&=0
            \end{alignedat}
\end{equation}
where for numerical feasibility the sums have to be suitably truncated. The expansion functions are either poloidal (``P'') or toroidal (``T''),
each of them representable by means of a  single spherical harmonic $\Ynm(\vartheta,\varphi)$:
\begin{align}
    &\begin{aligned}
    	   \Bvec_{nl}^{\tT m} &= \,- N_{nl}^{B\text{T}}\,	\ntoro{j_n(\mu^{B\text{T}}_{nl}\,r)\,\Ynm}\\*[2mm] 
    	   \Bvec_{nl}^{\tP m} &= \,- N_{nl}^{B\text{P}} \curl\,\ntoro{j_n(\mu^{B\text{P}}_{nl}\,r)\,\Ynm}  \\*[2mm]
    	   \Uvec_{nl}^{\tT m} &=\,- N_{nl}^{u\text{T}}  \,     \ntoro{j_n(\mu^{u\text{T}}_{nl}\,r)\,\Ynm} \\*[2mm] 
    	   \Uvec_{nl}^{\tP m} &= \,- N_{nl}^{u\text{P}} \curl\,\ntoro{\big(j_n(\mu^{u\text{P}}_{nl}\,r)-} \\
	                      &   \quad\quad\quad\quad\quad\quad\quad r^n j_n(\mu^{u\text{P}}_{nl})\big)\,\Ynm
    \end{aligned}\label{modes}\\
   & n=1,\ldots\;\; ; \quad m=0,\ldots,n\;\; ;\quad l=1,\ldots \quad .  \nonumber
\end{align}
Here, $j_n(x)$ is the $n$th order spherical  Bessel function of the first kind; the eigenvalues $\mu^{B\text{P}}_{nl}, \mu^{B\text{T}}_{nl}, \mu^{u\text{T}}_{nl}, \mu^{u\text{P}}_{nl}$, ordered increasingly with the index $l$,  are determined such that the boundary  conditions are obeyed. The collective index $i$ in \eqref{expan} is in \eqref{modes} resolved into $\{n,m,l,pt\}$, where $pt$ stands for  ``P'' or ``T''. The quantities $N$ are normalization factors.

The resulting system of ordinary differential equations for the mode amplitudes  $b_i(t)$ and $u_i(t)$ is integrated by use of
a Runge--Kutta scheme of 4th order with adaptive step--size control. 
Usually, the truncation of the expansion series ensured that the first neglected modes contained not more than 0.1 \% of the energy of the dominating one
and that the energy spectra showed a clear cutoff, mostly being exponential.   
	      
\section{Results}
\label{results}
\subsection{Solution characteristics}
Equations \eqref{ind}--\eqref{sfRB} allow of solutions with special symmetries with respect to the equatorial plane and the rotation axis, resp.:
By the symbol ``A'' we denote those solutions for which $\Bvec$/$\uvec$ is equatorially antisymmetric/symmetric. Analogously, ``S'' denotes solutions
with both fields being equatorially symmetric. 
Referring to the azimuthal dependence, the fields can either be both axisymmetric or superpositions of constituents
each characterized by a $\varphi$ dependence $\sim \exp(\mbox{i}m\varphi)$. In the general case all $m\ge0$ occur, but there are special solutions  for which
only constituents for a subset of $m$ values are non--zero. In the extreme case that this subset consists only of one $m$, the fields are denoted by this value or,
if there is an additional equatorial symmetry, by  ``A$m$'' or ``S$m$''.
For axisymmetric (with respect to the rotation axis) background fields the extreme case emerges just for the linearized versions of Eqs. \eqref{ind} and \eqref{mom}. That is,
each of the corresponding eigensolutions can be represented by a single $m$.
 
  In order to characterize the solutions more generally we use apart from energies and spectra the following quantities related to symmetry properties of the fields.
  The ``parity'' $\Pi$ reflects symmetry with respect to the equatorial plane: $\Pi=\pm1$ means perfect symmetry/antisymmetry. 
  The deviation from axisymmetry is described by the ``non--axisymmetry" $M$: $M=0,1$ refers to perfect axisymmetry and missing of any axisymmetric constituent, respectively. The general definitions are:
     \[
        \quad \Pi = \frac{E_\tsym - E_\tanti}{E_\tsym+E_\tanti}\,,\quad
        \quad M = 1-\frac{E^{0}}{E}
      \]
Here, $E_\tsym$, $E_\tanti$, $E^{0}$, and $E$ are the energies of the symmetric,
antisymmetric and axisymmetric ($m=0$) parts of a field and its total energy, respectively.

The runs were usually extended in time to the magnetic decay time, that
is, to $\ttdecay^B$. 
For the calculation of $\ttA$ we use the r.m.s. value of the magnetic field. In order to clarify the difference between the purely ohmic decay of a field and its evolution under the influence of the flows 
driven by its own Lorentz force, we relate the final magnetic energies of both cases after equal evolution times, where the final energy of the purely ohmic case is denoted by $\Emag^\tOhm$.  
Growth/decay times for a certain perturbation symmetry ($m$) extracted from the linear stages (see Tables \ref{decay_times} and \ref{growth_times}) are those of the fastest growing of all modes for this $m$. 

\subsection{Dipolar magnetostatic equilibrium}
In this model the strength of the initial background field, $B_0$, is chosen
such\footnote{For $\Pm=1$ and  as
$\ttAO/\min\{\ttdecay^B,\ttdecay^u\}=\sqrt{4\pi\rho} \cdot \eta
\max(\pi^2,\Pm\cdot15)$ the normalization field is then $10^{-3}\cdot \tilde{B}_{\text{r.m.s.}}(t=0)$ so that the initial dimensionless field is $10^3$.} that 
    \[
    \ttAO/\ttdecay =
     \begin{cases} 1.5\cdot10^{-2} \quad\text{for} \quad Pm=1,10\\
                    \phantom{.5}1    \cdot10^{-2} \quad\text{for} \quad Pm=0.1
     \end{cases} \]
As the symmetry parameters $\Pi$ and $M$ are defined with respect to the rotation axis, they depend on $\alpha$. For $\alpha = 0^\circ, 45^\circ, 90^\circ$ 
the initial values $\Pi=-1,0,1$ and $M=0,0.5,1$ are obtained, respectively. 
The results for $\Pm=0.1,1,10$ and the various $\alpha$ are summarized in \tablref{res_mageq}.
%
\subsubsection{Dependence on the rotation period}
For fixed $\Pm$ and $\alpha$ the final magnetic energy increases and the final kinetic energy decreases with decreasing $P$. Simultaneously the magnetic symmetry parameters   approach those of the initial field.
The three cases, in which the final magnetic energy is almost equal to the one resulting from ohmic decay only, are stable as confirmed
by examination of the linear stage. All other cases turn out to be unstable ones.
Note, that in case of extremely fast rotation ($q_P=0.012$ or $P=0.6$ ms) even background fields with $\alpha=90^\circ$ can be stabilized.
\subsubsection{Dependence on the magnetic inclination}
If we fix $q_P$=0.12,1.2 and $\Pm=1$, but consider $\alpha=45^\text{o}$ and $\alpha=90^\text{o}$
it turns out that at the same rotation rate and increasing $\alpha$  the stabilizing effect of rotation vanishes; for $q_P$=0.12,  $\alpha=45^\text{o}$ still instability is observed, but apparently 
close to its marginal value.
\subsubsection{Dependence on the magnetic Prandtl number}
Since both diffusivities are heavily overestimated in our calculations, we decided to adjust the normalization such that in comparison to the case $\Pm=1$ the case $\Pm=0.1$ corresponds to a reduction of the viscosity by a factor 10 while for $\Pm=10$ the same holds for the magnetic diffusivity.  
Thus we reduced in both cases the total dissipation in comparison with the case $\Pm=1$. Consequently, after the same elapsed period of time the total energy for $\Pm\neq1$ is always larger than for $\Pm=1$.
Reducing $\Pm$ from 1 to 0.1 obviously doesn't cause qualitative changes of the stability properties. 
An increase of $\Pm$, however, impedes stabilization, as visible from the cases with $\alpha=0$, $q_P=0.12$.
As $\Pm$  actually increases with progressing cooling, perhaps an even more rapid rotation is necessary to ensure
stabilization in the cooler stages. 
\begin{table*}
       \caption{\label{res_mageq} Results for the dipolar magnetostatic equilibrium model. 
       All quantities are taken after a period of $\ttdecay^B$ with some exceptions denoted in the 
       footnotes. Subscripts
       ``kin'' and ``mag'' refer to velocity and magnetic--field related quantities, respectively. Boldface symmetry symbols indicate, that the geometries of the
       initial and final magnetic configurations coincide thus giving a further hint for stability. The calculations for the non--rotating NS, $q_P=\infty$,
       were performed only for $\alpha=0$ because in this case the choice of the axis
       is of course arbitrary.  The results shown for shorter integration times, $\tau_\tend/\ttdecay^B<1$,  allow nevertheless unambiguous conclusions about stability.  } 
\centering  
     \begin{minipage}{\linewidth}
     \renewcommand{\thefootnote}{\thempfootnote} 
     \begin{center}   
      \begin{tabular}{cccccccccc}  
      \hline
      \hline\\*[-2.5mm] 
$\Pm$ & $\alpha\, (^\circ)$ & $q_P$  &  $\Emag/\Emag^\tOhm$ & $\Ekin/\Emag$&$\Pimag$ &	$\Pikin $ & $\Mmag  $&  $ \Mkin$& $\Bvec$ symmetry\\*[1mm]
       \hline\\*[-2.5mm]
  \multirow{8}{8mm}{0.1}& \multirow{4}{8mm}{ 0}  & $\infty$  &  0.0011 & 6.22    &  0.44 &  0.19 & 0.63 & 0.86  & mixed      \\
                        &			 & 12.       &  0.0024 &  1.93   & -0.44 & -0.15 & 0.17 & 0.85  & mixed      \\
                        &			 & 1.2       &  0.072  &  0.013  & -1.   &  0.66 & 0.049& 0.84  & $\sim $ A0 \\
                        &			 & 0.12      &  0.98   &  0.0003 & -1    &  1    & 0    & 0.9   & {\bf A0}   \\*[.5mm]
                                           \cline{2-10}\\*[-2.5mm]
                        & \multirow{1}{8mm}{ 45}
                         		         & 1.2       &  0.052  & 0.024   & -0.90 & 0.57  & 0.63 & 0.90  & mixed      \\*[.5mm]
                                           \cline{2-10}\\*[-2.5mm]
                        & \multirow{3}{8mm}{ 90} & 12.	     &  0.0014 & 6.51    &  0.009&  0.1  & 0.43 & 0.47  & mixed      \\ 
                        &			 & 1.2       &  0.0342 & 0.0033  &  0.85 &  0.4  & 0.8  & 0.97  & mixed      \\
                        &			 & 0.12      &  0.139  & 0.0009  &  1    & -0.38 & 1    & 0.94  &  S1   \\*[.5mm]
                                           \hline\\*[-2.5mm]
 \multirow{10}{8mm}{ 1} & \multirow{4}{8mm}{ 0}  &  $\infty$ &  0.0002 & 4.67    & -0.06 &  0.6  & 0.4  & 0.99  & mixed      \\
                        &			 & 12.       &  0.0004 & 6.6     & -0.5  &  0.8  & 0.2  & 0.88  & mixed      \\
                        &			 & 1.2       &  0.0076 & 0.0007  & -0.99 &  0.66 & 0.06 & 0.91  & mixed      \\
                        &			 & 0.12      &  0.98   & 0.00003 & -1    &  1    & 0    & 0.009 & {\bf A0}   \\*[.5mm]
                        		    \cline{2-10}\\*[-2.5mm]
                        &\multirow{2}{8mm}{ 45 } & 1.2       &  0.075  & 0.001   & -0.46 & -0.14 & 0.27 & 0.97  & mixed      \\
                        &			 & 0.12      &  0.824  & 0.00005 & -0.24 &  0.64 & 0.38 & 0.64  & mixed      \\*[.5mm]
                        		    \cline{2-10}\\*[-2.5mm]
                        &\multirow{4}{8mm}{ 90 } & 12.       &	0.0033 & 0.64	 &  0.5  &  0.78 & 0.77 & 1     & mixed      \\
                        &			 & 1.2       &  0.043  & 0.003   &  0.88 &  0.67 & 0.81 & 0.65  & mixed      \\
                        &			 & 0.12      &  0.14   & 0.00014 &  1	 &  1    & 1    & 0.59  & S1         \\
                        &                        & 0.012     &	0.98   & 0.0013  &  1    &  0.97 & 1    &  1    & {\bf S1}   \\*[.5mm]
                                           \hline\\*[-2.5mm]
  \multirow{5}{8mm}{ 10}& \multirow{3}{8mm}{ 0}  & $\infty$ &   0.00097& 0.305   & 0.155 & -0.512& 0.86 & 0.95  & mixed\\
                        &			 & 1.2      &   0.028  & 0.0041  & -0.40 & -0.32 & 0.41 & 0.91  & mixed\\
                        &		         & 0.12\footnote{$\ttend=0.72\ttdecay^B$}     &   0.77   & 0.0017  & -0.95  & 0.045 & 0.053& 0.94  & $\sim $ A0   \\*[.5mm]
                                           \cline{2-10}\\*[-2.5mm]

                        & \multirow{2}{8mm}{ 90 }
                         		         & 1.2      & 0.033   &0.0033  &0.19   & 0.23 & 0.68 &0.93    & mixed \\
                        &			 & 0.12\footnote{$\ttend=0.3\ttdecay^B$} & 0.74 & 0.0017 & 0.99 & 0.82 & 0.997 & 0.906 & $\sim $ A0   \\*[.5mm]
                                           \hline
                                         
       \end{tabular}
       \end{center}
       \end{minipage}
      \end{table*}    
\subsubsection{A stabilized case}
\begin{figure*}[t]  
  \begin{tabular}{@{\hspace{0.5cm}}c@{\hspace{-.1cm}}c@{\hspace{3mm}}}
 \multicolumn{2}{c}{\includegraphics[width=.85\textwidth]{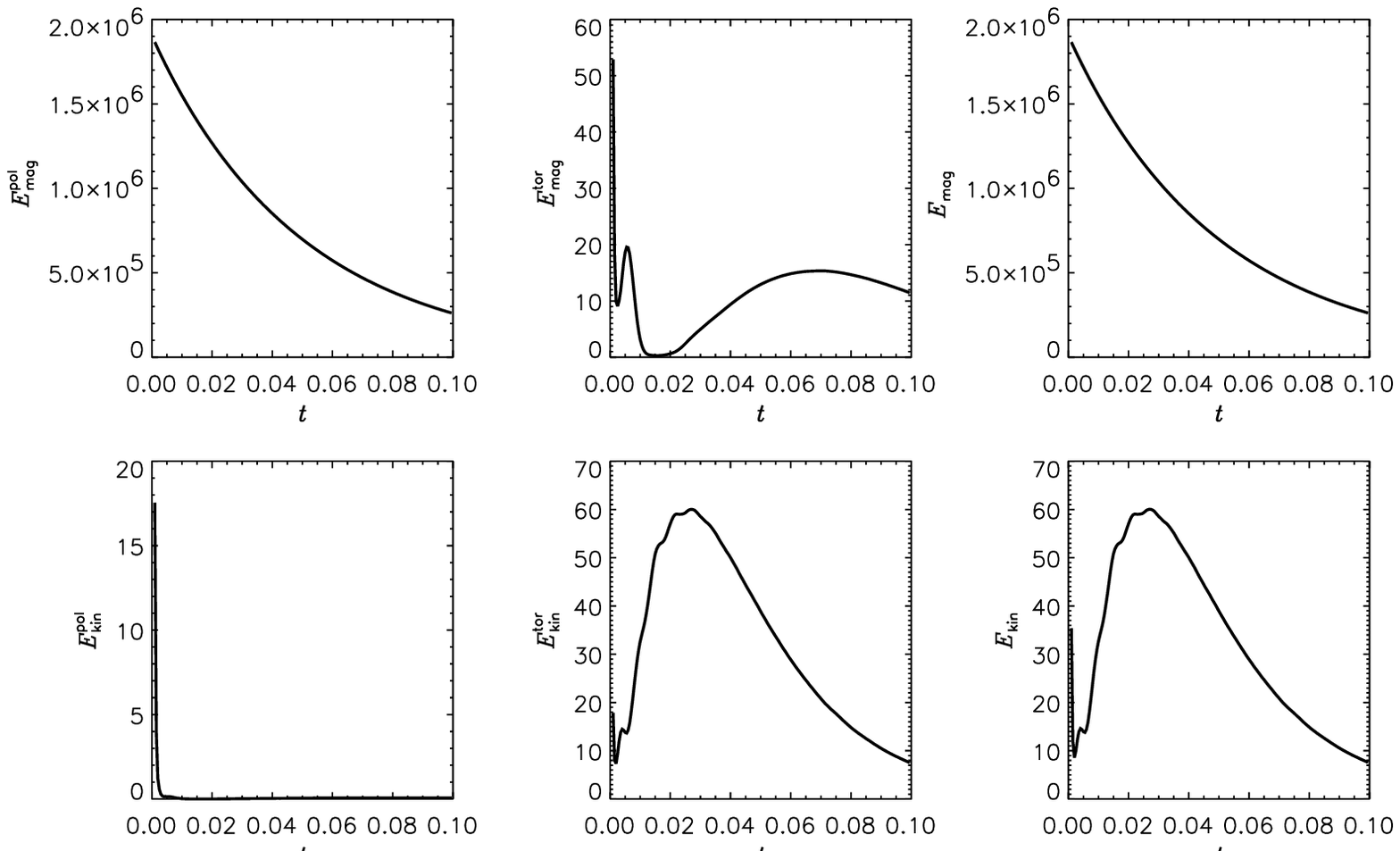}}  \\
\includegraphics[width=.45\textwidth]{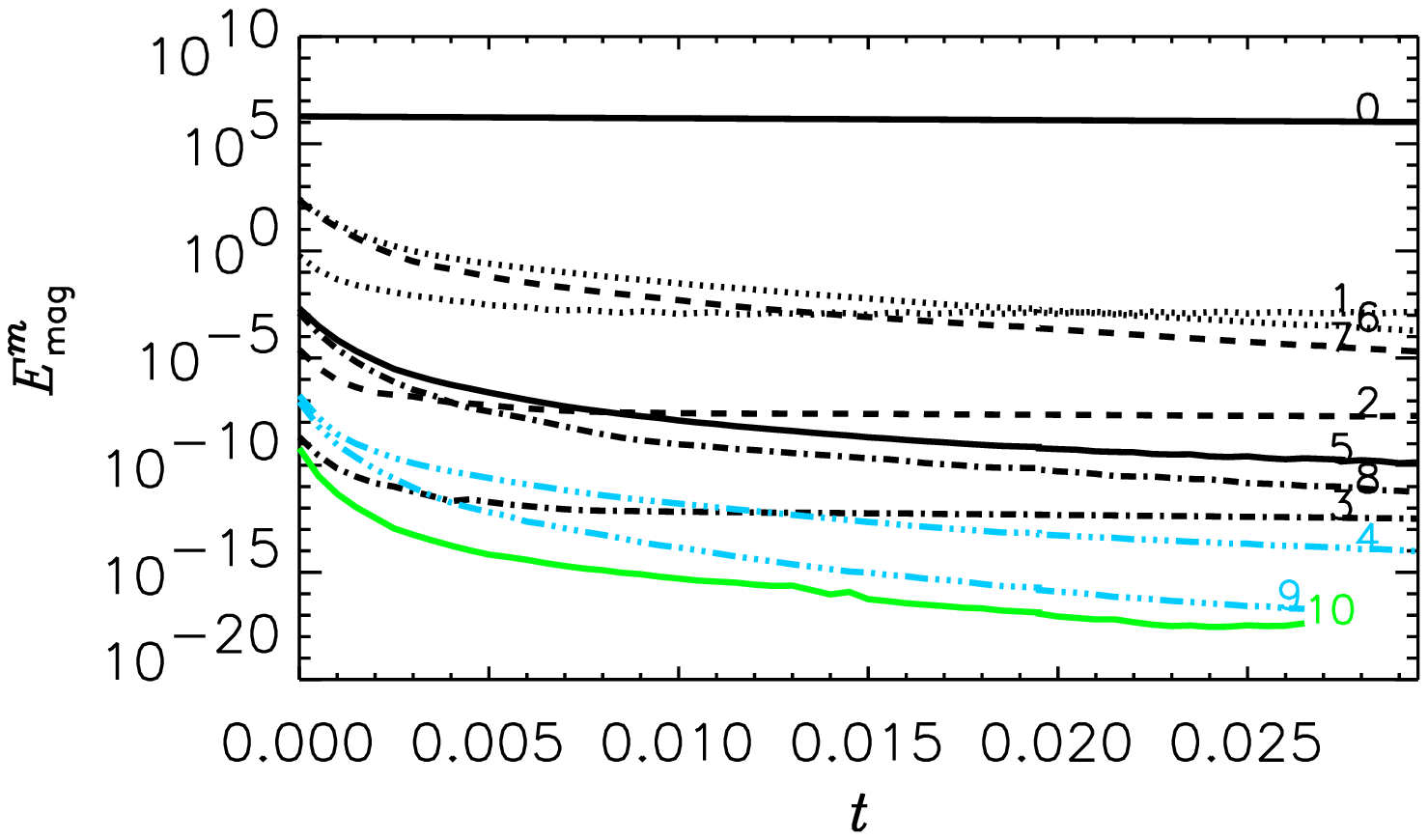} &       
\includegraphics[width=.45\textwidth]{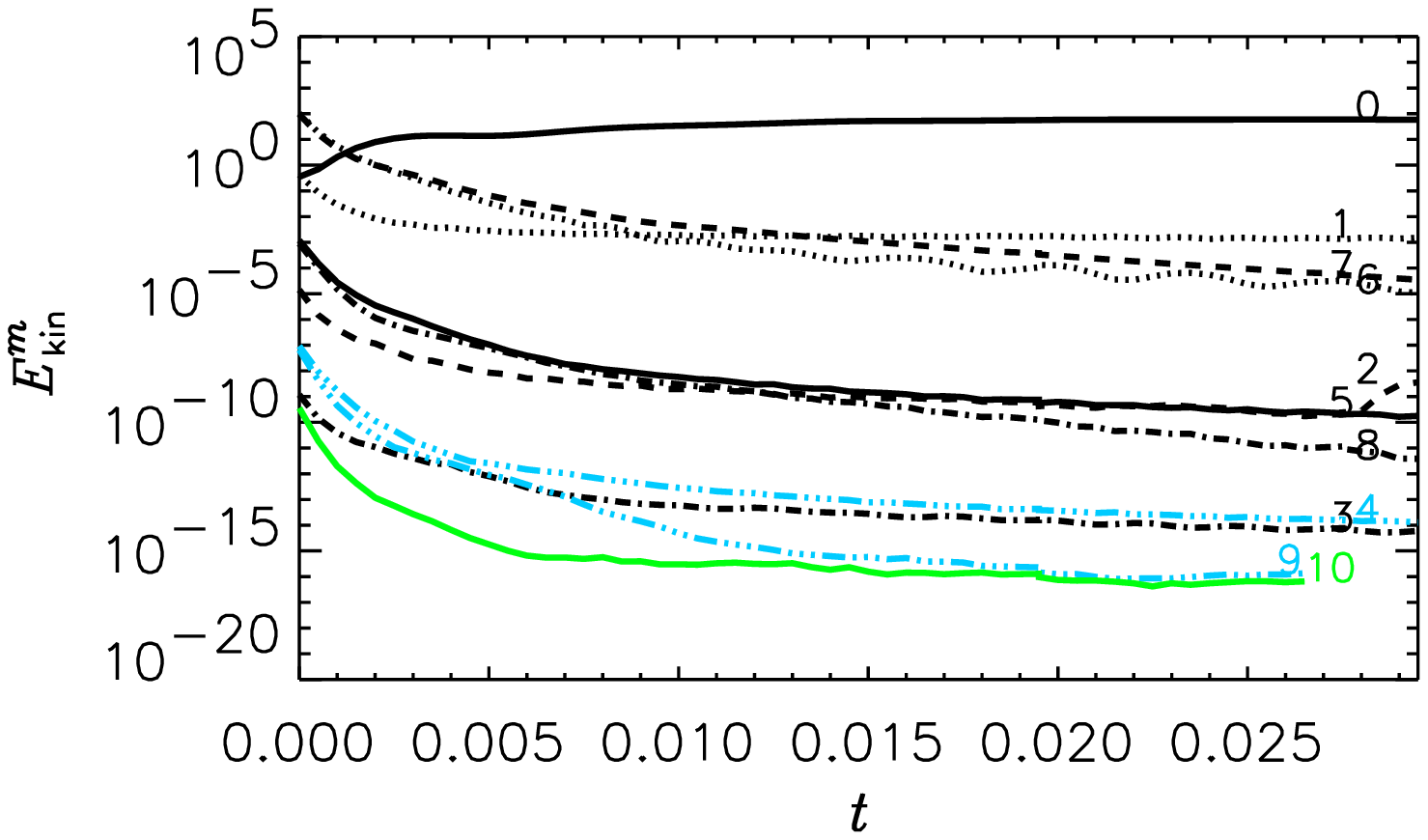}\\*[-4mm]
\includegraphics[width=.45\textwidth]{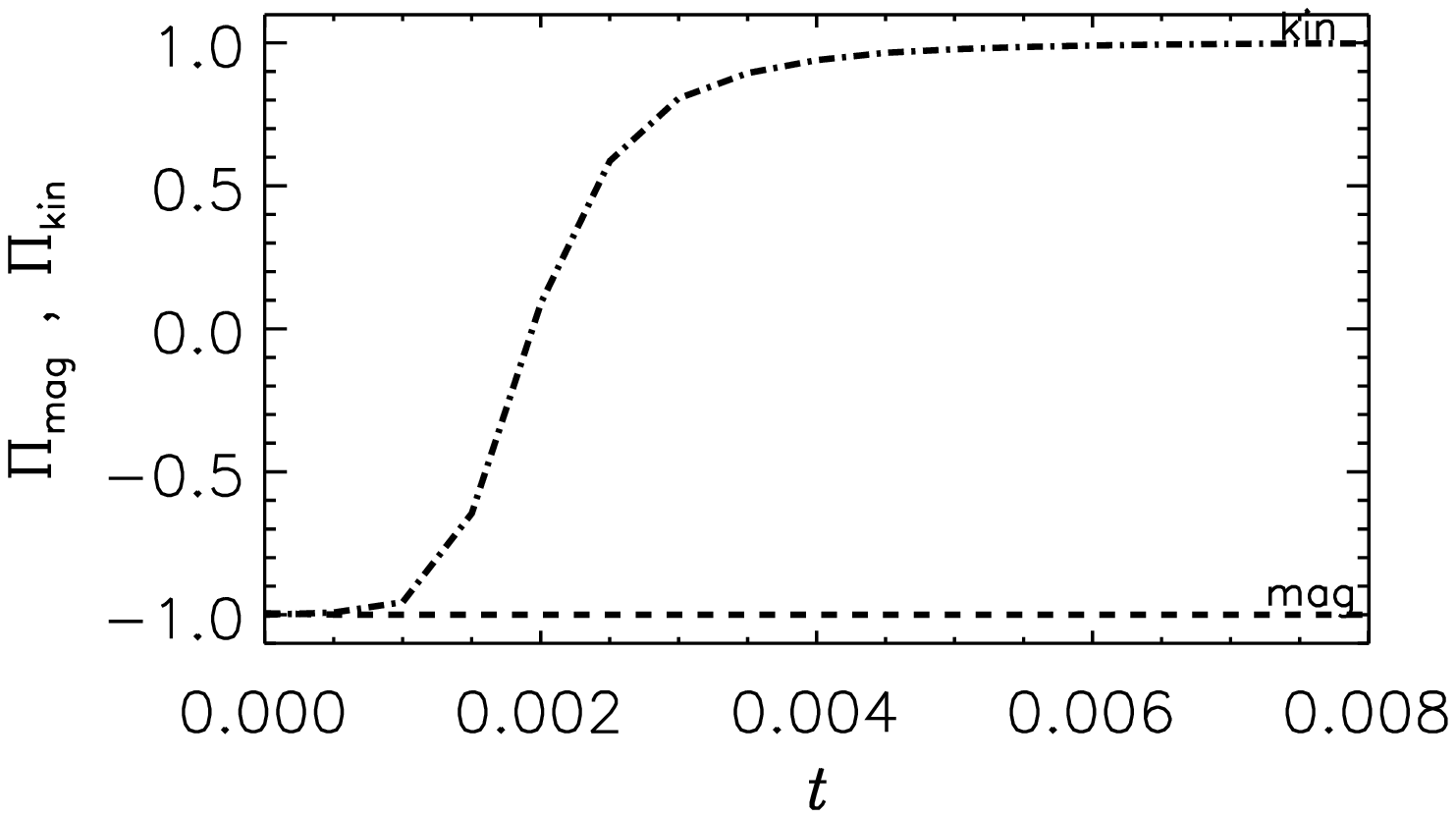} & 
\includegraphics[width=.45\textwidth]{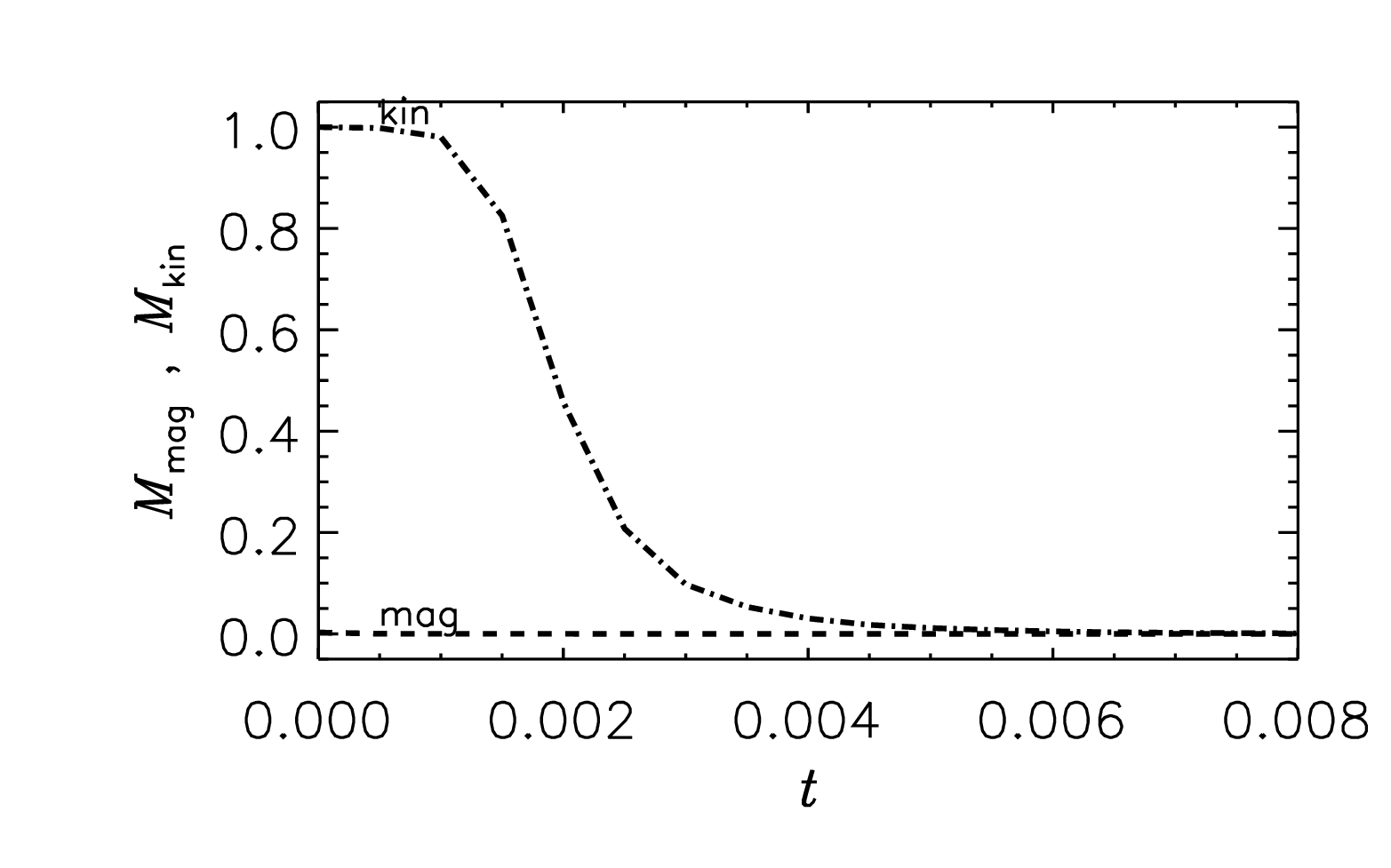} 
   \end{tabular}
   \caption{\label{mageq_stab1} Temporal evolution of the disturbed dipolar magnetostatic equilibrium for $q_P$=0.12, $\Pm=1$, $\alpha=0$.  Time is in units of $\tau_\tOhm$, energy in units of $\rho\eta^2/R^2$. Subscripts
       ``kin'' and ``mag'' refer to velocity and magnetic--field related quantities, respectively.  Upper two rows:  magnetic and kinetic energies, each further subdivided in its poloidal and toroidal part;  third row: $m$ spectra of the magnetic and velocity fields, $\Emag^m$, $\Ekin^m$, respectively. Lower left and right panels: parity and non-axisymmetry. 
}
 \end{figure*}

Let us consider the case $q_P=0.12$, $\Pm=1$, $\alpha=0$ which for the model parameters \eqref{denorm} corresponds to $P=6$ ms,  that is, a very fast spinning new--born NS. 
 Of course, the non--linear evolution is in general influenced by the arbitrariness of the initial conditions. However,
in the present case it is without relevance since all perturbations, apart from a short initial ``switch--on'', decay very fast
as shown in \figref{mageq_stab1}. There, time--changing energy spectra $E_\tmag^m(t)$, $E_\tkin^m(t)$ are shown in which all modes with the same $m$ (cf. \sectref{method}) are summed up to yield a single spectral contribution labeled by $m$.  Note, that the temporary growth of the kinetic energy
for  $m=0$
can be attributed to the error introduced by representing the background equilibrium as a truncated series of modes.

Due to the axisymmetry of the background field and all coefficients in \eqref{ind} and \eqref{mom}, each eigenmode of the linearized problem is with respect to
its dependence on $\varphi$ uniquely characterized by a single $m$. 
Therefore, our spectral representation with respect to $m$ allows identifying the decay time of the slowest
decaying eigenmode for each $m$ from the data of the linear stage. 
When performing this, one has to skip an initial phase in which the decay is
not purely exponential. This is a consequence of the
arbitrary initial conditions for the perturbations feeding several eigenmodes with different decay times simultaneously. However,
there is always an instant of time after which the slowest decaying mode survives as the only one yielding a single, well-defined 
decay time.

The decay times are shown in \tablref{decay_times} in units of $\ttAO$.
The value for $m=1$ is not included as apparently the slowest decaying mode belonging to $m=1$ is not yet clearly dominating over the faster decaying 
$m=1$ modes.

Moreover, \figref{mageq_stab1} shows the field parities and non--axisymmetries as functions of time. Clearly, the magnetic field retains its symmetry completely whereas the velocity starts with equatorial antisymmetry and complete non--axisymmetry as can be explained from the interaction of the perturbations with the background field, but
approaches very fast its final equatorially symmetric, axisymmetric geometry.
 Note, that even the maximum of  $\Ekin$ is much smaller than $\Emag$.
A comparison of the initial and the final state
reveals that the final state coincides almost exactly with 
the one resulting from free decay of the initial state alone. 
Hence, in all we can safely conclude that the considered dipolar equilibrium is indeed stabilized by rotation.
\begin{table}[htbp]
   \caption{
   \label{decay_times} Decay times $\tdec$ of the slowest  decaying eigenmodes depending on $m$ for the dipolar equilibrium model with $q_P=0.12$,  $\alpha=0$ and $\Pm=1$ (a stabilized case).}
   \centering
   \begin{tabular}{@{}>{$}c<{$}@{\hspace{4mm}}>{$}c<{$}@{\hspace{4mm}} >{$}c<{$}@{\hspace{4mm}} >{$}c<{$}@{\hspace{4mm}}>{$}c<{$}@{\hspace{4mm}}>{$}c<{$}@{\hspace{4mm}}>{$}c<{$}@{\hspace{4mm}}>{$}c<{$}@{\hspace{4mm}}>{$} c<{$}@{}} 
   \hline
   \hline\\*[-2.5mm]
   m   &  2 & 3 & 4 & 5 & 6 & 7 & 8 \\*[.75mm]
    \hline\\*[-2.25mm]
\frac{\tdec}{\ttAO}  & 40.97  & 24.73 & 20.82 &  13.94 & 10.89  &  8.50 & 8.31           \\
    \end{tabular}
\end{table}

\subsubsection{An unstable case}
It is instructive to compare the former case with a much slower rotating NS: $P/\ttAO=12$, i.e. the same setup as before, but $P=0.6$ s.
\begin{figure*}[h]
%
\begin{tabular}{@{\hspace{.5cm}}c@{\hspace{-.1cm}}c@{\hspace{3mm}}}
   
\multicolumn{2}{c}{\hspace*{-.2cm}\includegraphics[width=0.8\textwidth]{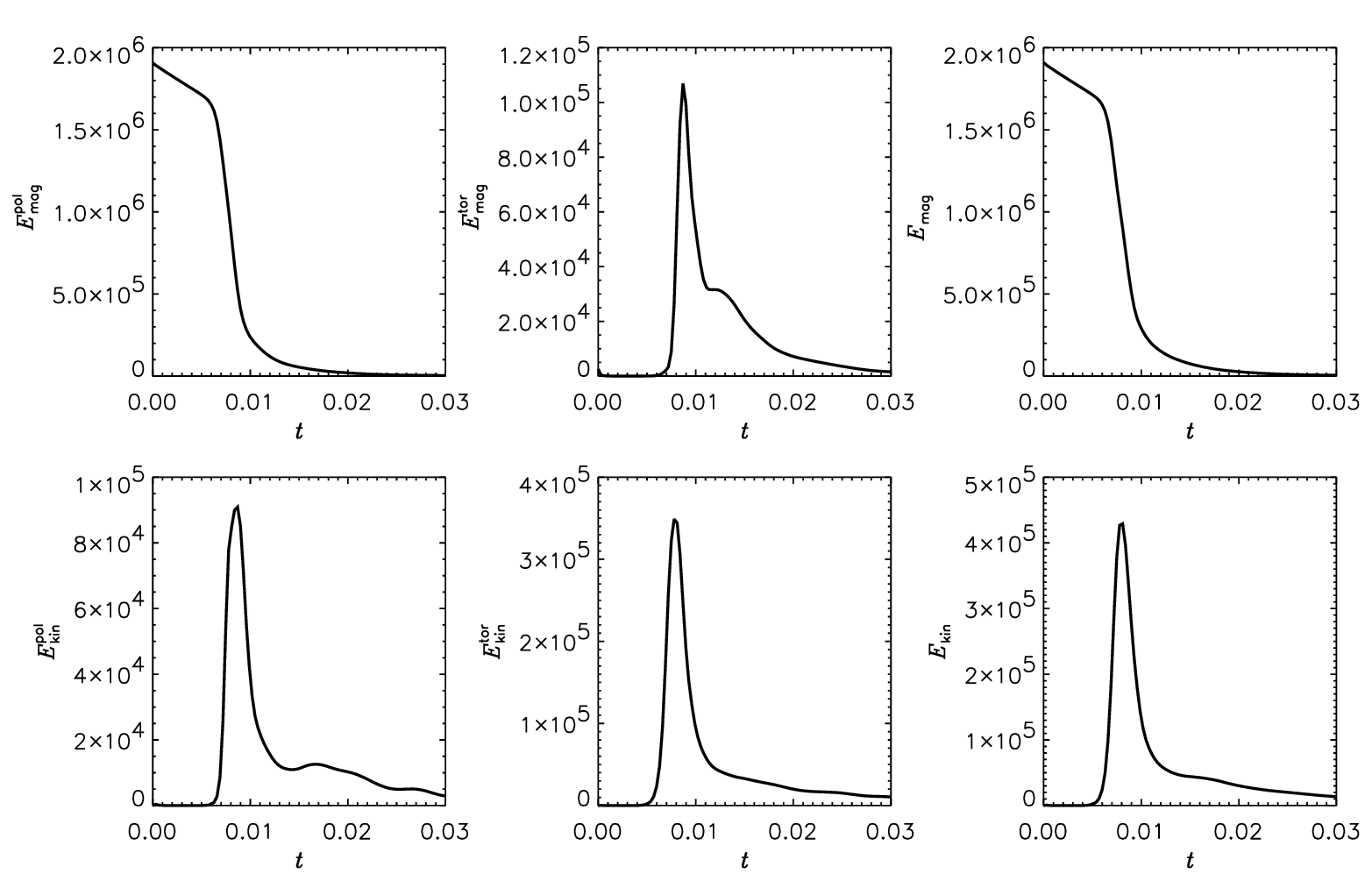}}\\
    \includegraphics[width=.45\textwidth]{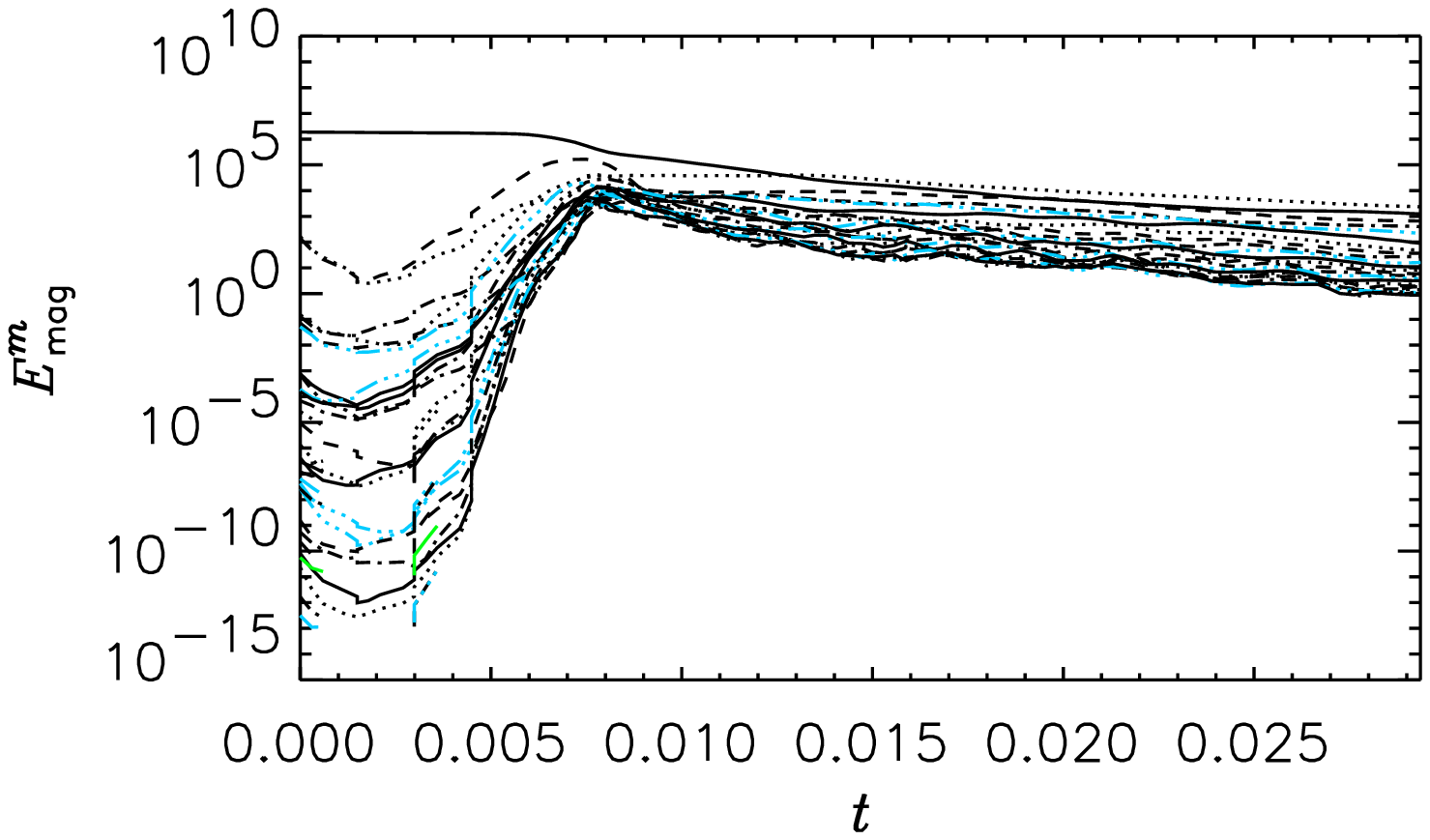}&
    \includegraphics[width=.45\textwidth]{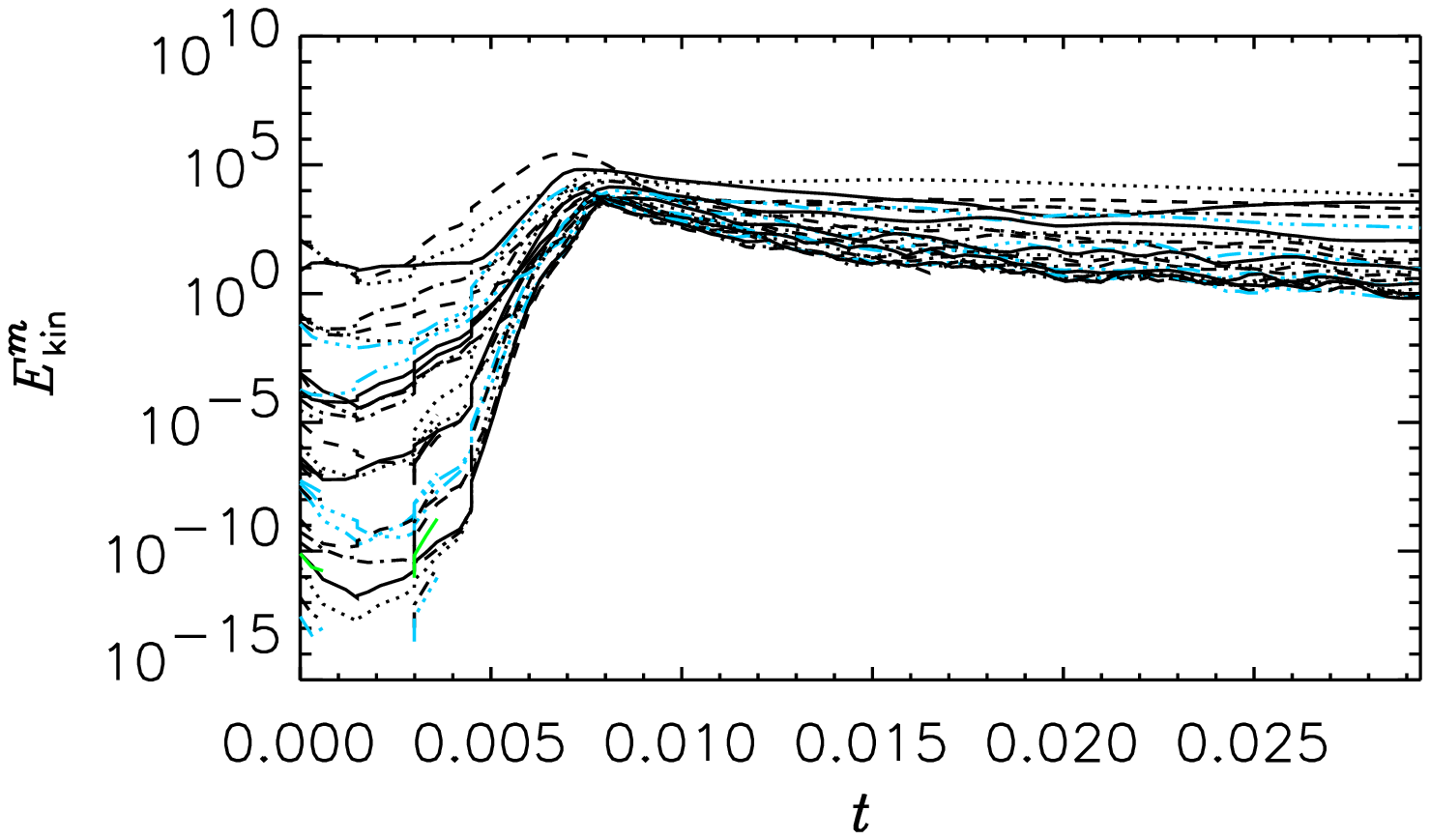}\\*[-4mm]
    \includegraphics[width=.45\textwidth]{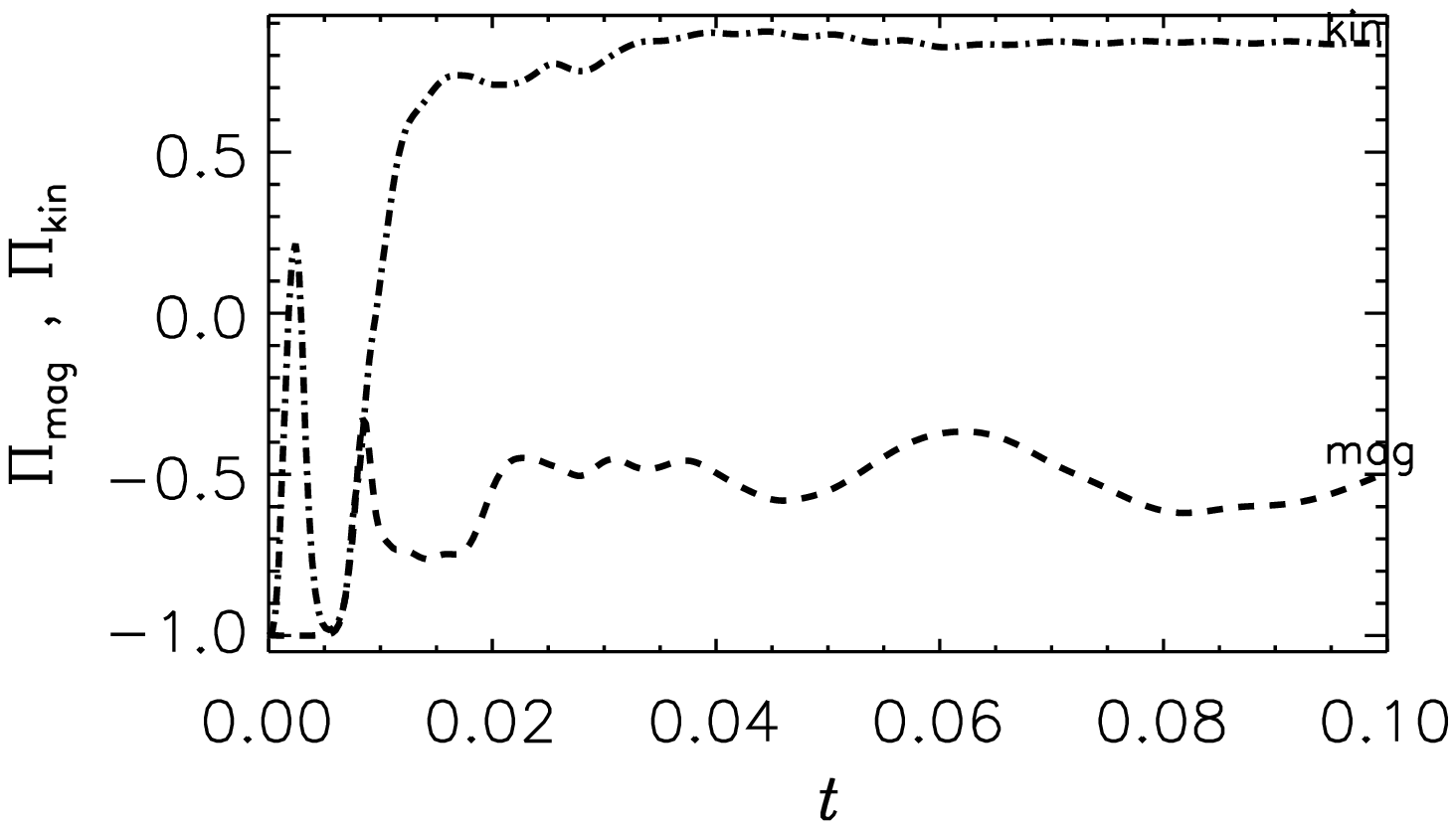}&
    \includegraphics[width=.45\textwidth]{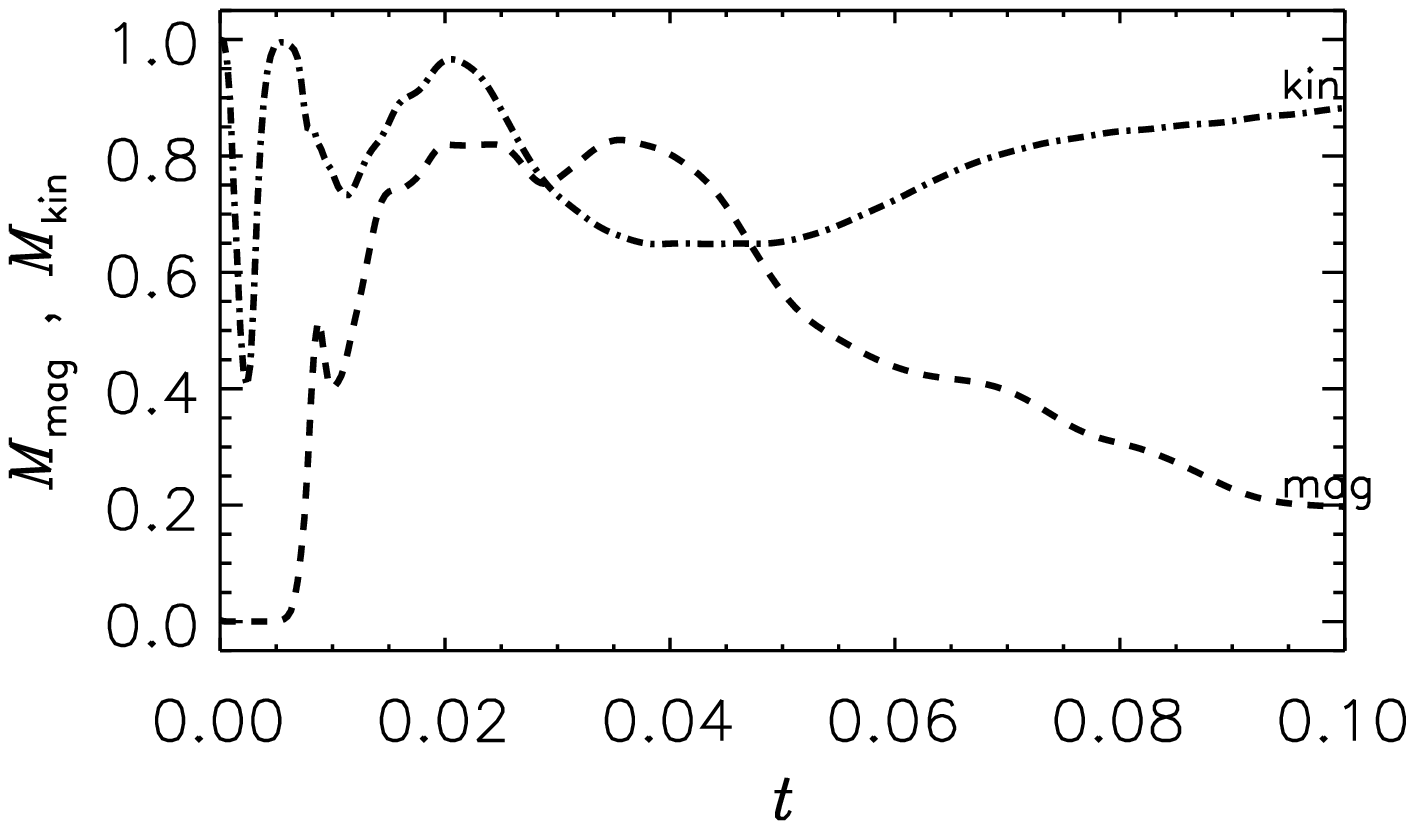}
  \end{tabular}
  \caption{\label{mageq_instab}Temporal evolution of the disturbed dipolar magnetostatic equilibrium. $q_P$=12, $\Pm=1$, $\alpha=0$. For further explanations see \figref{mageq_stab1}.
   Note that the numerical resolution with respect to $m$ in the vicinity of the kinetic energy
   peak, though being worse than the usually obeyed standard (cf. end of Sect.
   \ref{method}), still ensures an exponential spectral decay of $\sim$ two orders of
   magnitude. Using the parameters \eqref{denorm} $t=0.01$ corresponds to 0.5 seconds!}
%
\end{figure*}
\begin{figure*}[t]  
%
  \begin{tabular}{@{\hspace{1cm}}c@{\hspace{-1.cm}}c@{\hspace{3mm}}}
\includegraphics[width=.5\textwidth]{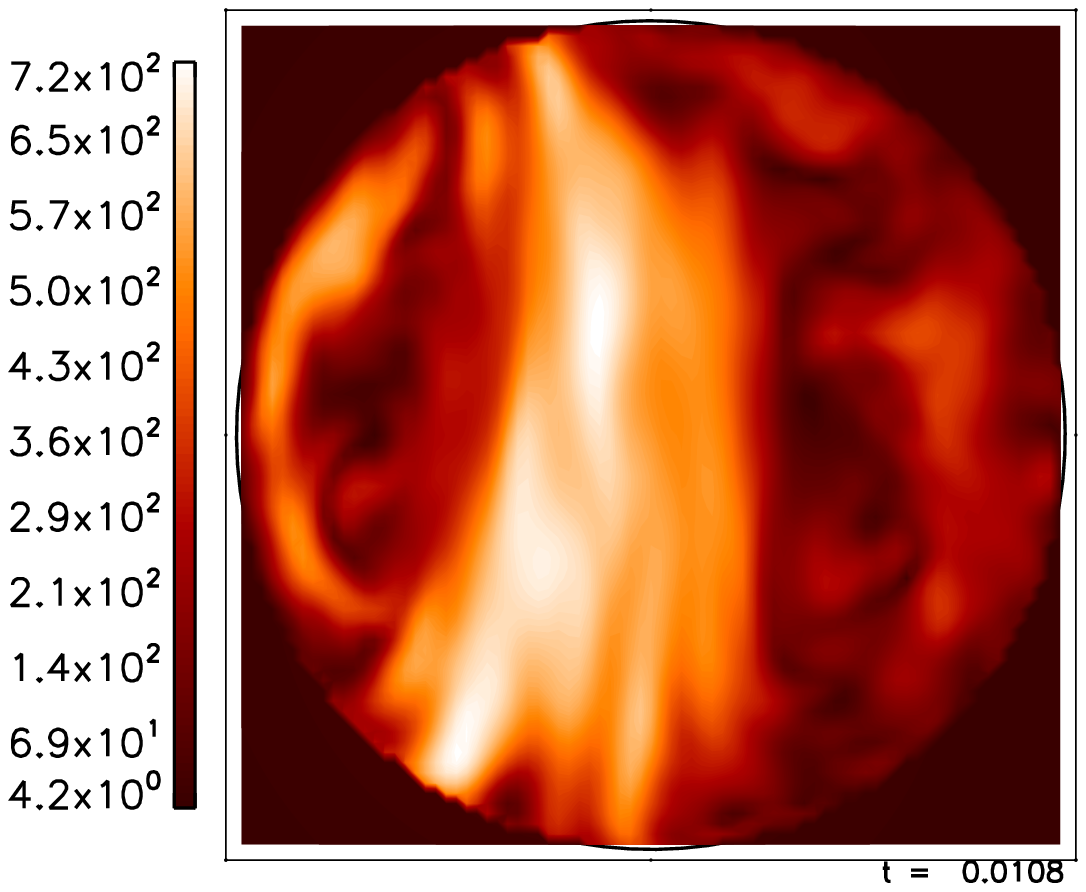}\hspace{-.5\textwidth}\includegraphics[width=.5\textwidth]{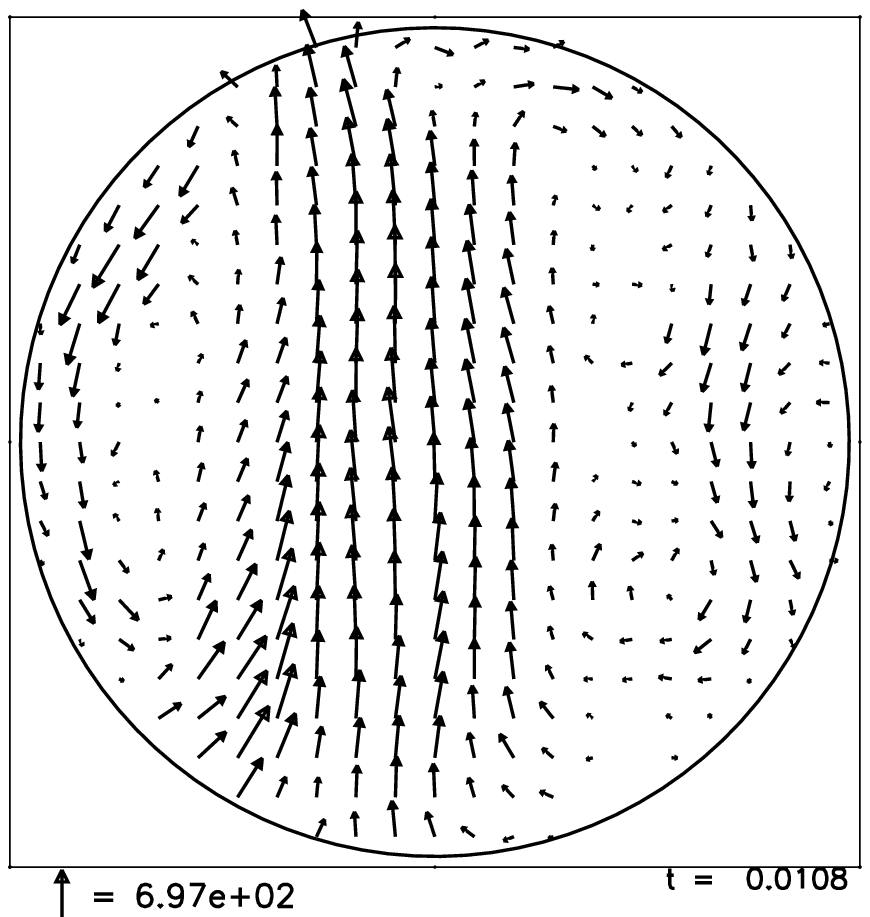} &       
\includegraphics[width=.5\textwidth]{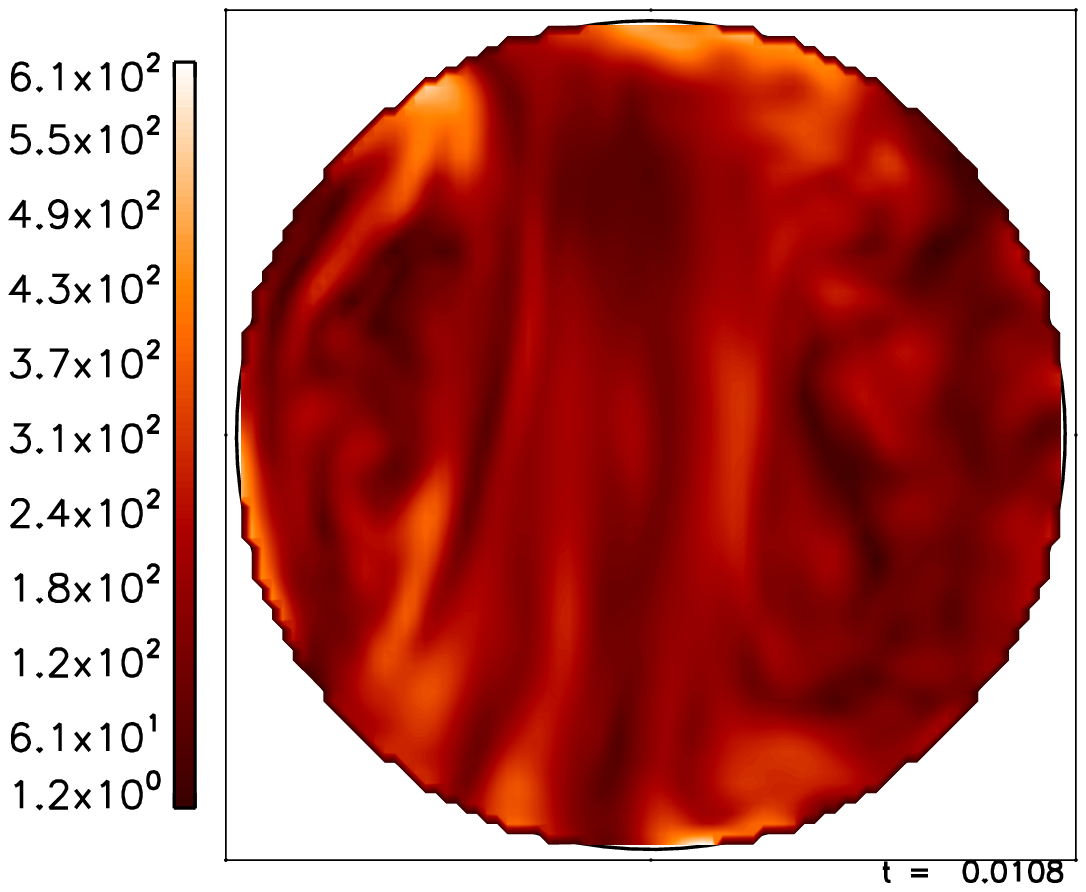}\hspace{-.5\textwidth}\includegraphics[width=.5\textwidth]{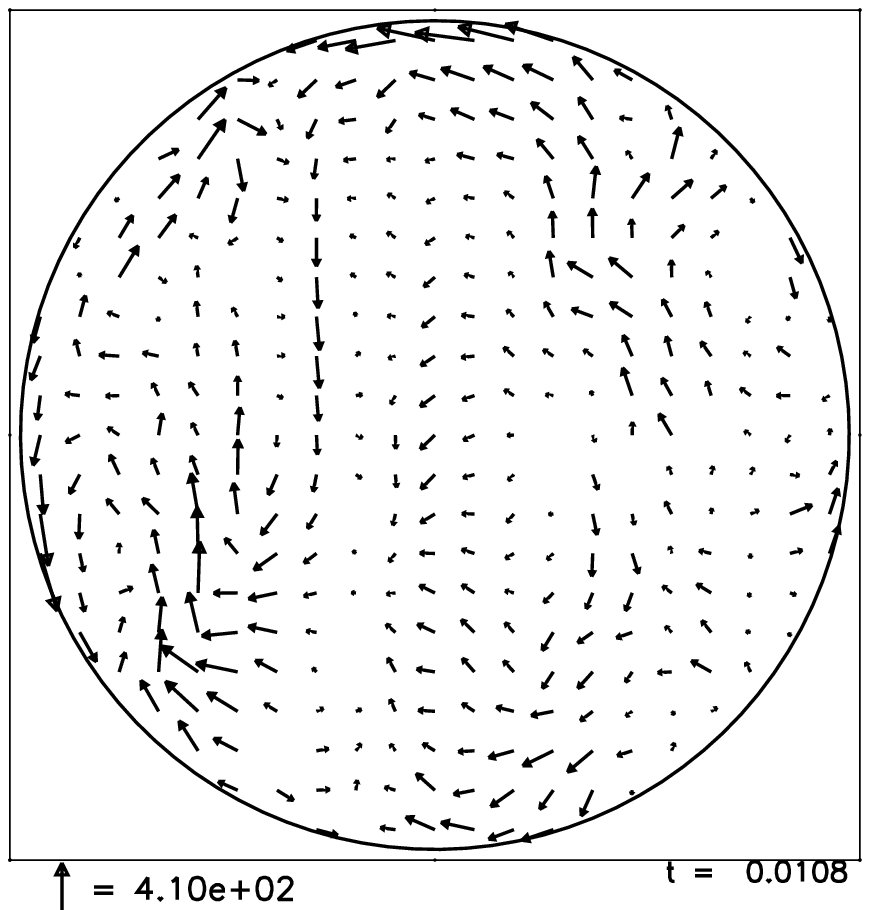}
   \end{tabular}
   \caption{\label{mageq_instab2}  Field geometries of magnetic field (left) and flow (right) around the end of the dramatic field reduction phase (see upper panels of {\figref{mageq_instab}} at $t\approx 0.01$) in  a meridional plane for the dipolar equilibrium model with $q_P$=12, $\Pm=1$, $\alpha=0$. Arrows
   indicate vector components parallel to the paper plane. Their maxima are
   $6.93\cdot 10 ^{14}$ G and $8.3\cdot10^{6}$ cm s$^{-1}$, respectively. Colors encode the field moduli: the brightest tone corresponds to $7.2\cdot 10^{14}$ G  and  $1.2\cdot10^{7}$ cm s$^{-1}$, respectively. }
 \end{figure*}  

Inspecting \figref{mageq_instab}, it becomes clear that the slower rotation is no longer capable of stabilizing the background field.
The kinetic energy of the evolving  flow is in its maximum about $10^4$ times larger than in the stable case with rapid rotation. 
In the temporal evolution of the 
energies we can distinguish three different phases: initially, up to $\approx6\ttAO$, the effect of the perturbations is negligible and the slow ohmic decay dominates. 
Afterwards, the rapidly increasing perturbations drain out energy from the background field very efficiently. From $\approx8\ttAO$ on, saturation sets in followed by a 
decelerating decay of all energy constituents to low values.

Checking the validity of the incompressibility assumption, we find that, except for a very thin outer shell where in real NSs the density would be less than $10^8$ g cm$^{-3}$,
the maximum velocities are well below the speed of sound. However, we have to concede that for realistic viscosities 
this assumption is likely to be violated during the fully developed non--linear stage, at least locally. A compressible treatment would then be necessary.
On the other hand, during the early phase of the non--linear stage, that is, before the kinetic energy peaks, the observed time scale is short 
enough to be unaffected by the assumed unrealistically high diffusivities. Therefore, we are confident that up to this moment our results are reliable with respect
to magnitudes, too. 

The time--changing $m$--spectra in \figref{mageq_instab}
show nearly exponential growth of the perturbations on the linear stage from which growth times $\tgrow$
were deduced  (see \tablref{growth_times}). While for low $m$ the behavior of
$\tgrow/\ttAO$ with respect to $m$ is non--monotonous, a continuous decrease is observed beyond $m=6$. This is in accordance with predictions from analytical work \citep{PT85}.

The time dependences of parity and non--axisymmetry shown in \figref{mageq_instab} demonstrate the increasing deviation from the equilibrium, too.
\begin{table}[h]
   \caption{
   \label{growth_times} Growth times $\tgrow$ of the fastest  growing eigenmodes depending on $m$ for the dipolar equilibrium model with $q_P=12$,  $\alpha=0$ and $\Pm=1$ (an unstable case).}
   \centering
   \begin{tabular}{@{}>{$}c<{$}@{\hspace{3mm}}>{$} c<{$}@{\hspace{3mm}} >{$} c<{$}@{\hspace{3mm}}>{$} c<{$}@{\hspace{3mm}}>{$} c<{$}@{\hspace{3mm}} >{$} c<{$}@{\hspace{3mm}} >{$} c<{$}@{\hspace{3mm}}>{$} c<{$}@{\hspace{3mm}} >{$} c<{$}@{} } 
   \hline
   \hline\\*[-2.5mm]
   m   & 1&  2 & 3 & 4 & 5 & 6 & 7 & 8 \\*[.75mm]
    \hline\\*[-2.25mm]
    \frac{\tgrow}{\ttAO}   &  0.45  &  0.25  &   0.17  &  0.23  &  0.48  &  0.90   &   0.79 &    0.32   
    \end{tabular}
\end{table}
  
\subsection{Internal uniform field}
Here, $B_0$  is chosen such that 
    \[
    \ttAO/\ttdecay = 8\cdot10^{-3} \]
and its symmetry parameters $\Pi$ and $M$ coincide with those of the dipolar equilibrium field. The results for $\Pm=1$ and $\alpha = 0^\circ, 45^\circ, 
90^\circ$ are summarized in \tablref{res_florud}. The general tendencies are the same as for the dipolar equilibrium field. Again, the significant influence of
$\alpha$ on the stability is proven. 
\begin{table*}  
\begin{minipage}{\linewidth}
\renewcommand{\thefootnote}{\thempfootnote}

\begin{center}
       \caption{\label{res_florud} Results for the internal uniform field model. Only $\Pm=1$ was considered. For explanations see Table \ref{res_mageq}.
       Boldface symmetry symbols denote here that the final geometry coincides with
       the one of the dipolar equilibrium. }
      \begin{tabular}{ccccccccc}      
         \hline       
         \hline\\*[-2.5mm]     
$\alpha\, (^\circ)$ & $q_P$  &  $\Emag/\Emag^\tOhm$ & $\Ekin/\Emag$ & $\Pimag$ &	$\Pikin$  & $\Mmag$  &   $\Mkin$ &  $\Bvec$ symmetry\\*[1mm]
       \hline\\*[-2.5mm]
                        \multirow{4}{8mm}{ 0}  &  $\infty$  & 0.00009 &  147.6   & -0.41  & -0.43 &  0.82  & 0.81   & mixed\\
                        		       & 12.      & 0.001   &  2.92    & -0.82  & -0.44 &  0.11  & 0.86   & mixed\\
                        		       & 1.2      & 0.083   &  0.0034  & -1	&  0.22 &  0.03  & 0.93   & mixed\\
                        		       & 0.12     & 1.02    &  0.00002 & -1	&  1	&  0	 & 0.0001 & {\bf A0}   \\*[.5mm]
                        		    \hline\\*[-2.5mm]
                        \multirow{2}{8mm}{ 45 }
                        		       & 0.12     & 0.56    &  0.00006 & -0.33  & -0.15 &  0.34  & 0.92   & mixed \\
                        		       & 0.012    & 1.006   &  0.0023  & -0.011 & -1    &  0.49  & 1      & {\bf A0+S1} \\*[.5mm]
                        		    \hline\\*[-2.5mm]
                        \multirow{2}{8mm}{ 90 }
                        		       &  0.12    & 0.193   &  0.0005  & 1  & 1 &  1 &   0.91   & S1 \\
                        		       &  0.012   & 0.976   &  0.0013  & 1  & 1 &  1 &   0.997  &  {\bf S1} \\*[.5mm]
                        		    \hline
      \end{tabular}
    \end{center}
  \end{minipage}
\end{table*}
In the cases $\alpha=0$, $q_P=0.12$ and $\alpha=45^\circ$, $q_P=0.012$ the final magnetic energy is slightly larger in the MHD case in comparison with the 
purely ohmic decay. This somewhat surprising result is not in contradiction to energy conservation because it cannot be excluded that the flow driven by the Lorentz force reorganizes the
magnetic field in such a way that small scale components just below the surface (cf. \sectref{init_florud}) are smoothed out more effectively than by ohmic diffusion only, thereby decelerating the overall decay.
 
\begin{figure}[htbp]
 
  \vspace{-5mm}
   \hspace{-.1cm}\includegraphics[width=.5\textwidth]{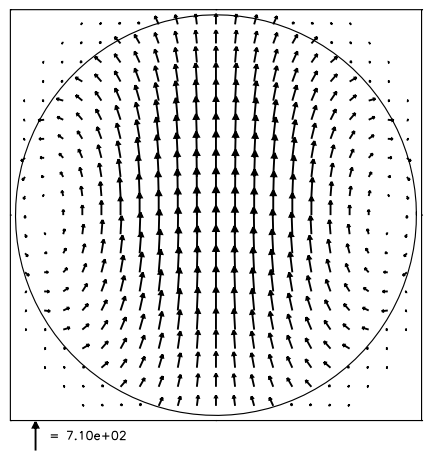} 
    \caption{ \label{pol_final} Final field geometry for the internal uniform field model in a meridional plane (the field is almost exactly axisymmetric) for $\alpha=0$, $q_P=0.12$, $\Pm=1$. With the denormalization based on Eq. \eqref{denorm} the maximum field  strength is about $7.1\cdot10^{14}$ G.}
\end{figure}

A comparison of both the final spectra and field geometry  of the result for  $\alpha=0$, $q_P=0.12$ with the corresponding result for the 
dipolar equilibrium model gives strong evidence for
a tendency to approach the same state for $t\rightarrow\infty$ (see \figref{pol_final}). Not only the difference in the initial geometry (cf. Figs. \ref{uniform} and \ref{poleq}), but also the difference in the initial energies is
obviously equalized after having gone through the non--linear stage. This is confirmed by the similarity of the spectra of the two results the remaining differences of which disappear
in the further course of time. \figref{pol_final} shows the final field for the internal uniform field model which is almost identical with the final field
of the poloidal equilibrium model. The relative r.m.s. value of the difference of both fields is only 1.6 \%.
The same coincidence is found for $\alpha=45^\circ,90^\circ$, $q_P=0.012$.

\section{Discussion and Conclusions}
\label{disc}
The results of the present study suggest the following major conclusion:
Never again in an isolated NS's life its magnetic field may suffer  so drastic changes as during the first at most $1000$ seconds after birth, provided the dipolar field present immediately after the proto--NS phase has magnetar strength.
Depending on the rotational velocity and the inclination of the field with respect to the rotation axis,
the inborn field will either survive that period as a magnetar field or lose almost all of its initial magnetic energy by transferring it into magnetic and kinetic energy of quite small--scaled fields  eventually being consumed by ohmic and viscous dissipation.
After the end of the unstable phase, large--scaled magnetic fields dominate again, but have
a much lower strength.    

We found that purely dipolar fields in highly magnetized new--born NSs are
unstable on a time--scale of a  few \Alf{} times unless fast rotation, i.e.,
$\Omega_\tAlfven/\Omega \la 0.1$, acts stabilizing. The latter condition,
however, is only a necessary one, fulfilled for NSs which have at the end of
their proto--NS phase a poloidal field strength of $\sim 10^{15}$G  and a rotation period of $1\dots 6$ ms. 

As a sufficient condition, the above presented results yield that rotation is only efficiently stabilizing as long as the inclination angle between rotation and magnetic axis $\alpha <  45^\circ$.
Therefore, we may speculate that those NSs whose inborn field is not too much bent away from the rotation axis and which rotate with a initial period $\la 6$ms will continue their life as magnetars, while for quite similar NSs, which, however, are born as orthogonal rotators, even rotation periods  $\approx 6$ms  will not stabilize the poloidal field against its substantial reduction down to standard pulsar strengths.
As another conclusion we suggest that for each inclination angle $\alpha$ there exists a ``final'', apart from ohmic decay, stable field configuration, whatever the inborn field shape has been.
 Indeed, they seem to be identical up to a rotation by $\alpha$.

\citet{FR77} consider qualitatively the relaxation of a uniform field into its lowest energetic state. Assuming  a relatively large \Alf{} time scale of $\sim 100$ s (in 1977 magnetar field strengths were  inconceivable!) and a rotation period of the new--born NS $\approx\!6$  ms,  they conclude that the time for the readjustment of that field is about 10 days, clearly beyond the onset of crystallization which impedes the relaxation already after a few 1000 seconds. Their relaxation time scale is (except for a factor of 2) that of the Tayler instability, $\ttA$,  enhanced by a factor $\ttA/P$ due to the strong effect of the Coriolis force \citep{S99}. 
Making use of this argumentation with respect to magnetar field strengths, however, this estimate for the relaxation time scale reduces to $\sim 1$ s, i.e.,  the initially uniform magnetic field would have ample time to establish its energetically lowest state which is supposed to be no longer dominated by a large scale field.
In contrast,  our study shows that this will not happen for a rotation period of $\lesssim 6$ ms
unless $\alpha\ga 45 ^\circ$.
For an understanding of this discrepancy the behavior of the so--called Levitron\textsuperscript{TM} can be instructive \cite[cf. e.g.][] {JWG97}: Similar to the setup of \citet{FR77} it consists of two permanent, mutually repelling magnets where a stable (levitation)  situation can be achieved only via rotation.
However, it seems questionable at all whether the analogy with permanent magnets  is appropriate for NSs because the latter can be described in terms of ideal MHD characterized by a frozen--in field whereas the former are characterized by frozen--in currents.

In this paper we consider the stability of purely poloidal fields. However, it is very unlikely, that in nature such a configuration exists. More likely is that the poloidal field is accompanied by a toroidal one, which, if it is of comparable strength, may act stabilizing as well \citep{T80,BS04,BN05}. The dependence of the stability of such a configuration  on rotation and inclination will be studied in a forthcoming paper.

It is well--known, that the effect of a NS's stably stratified density profile exerts
another stabilizing effect \citep[see, e.g.,] []{PT85}.
This is due to the strong suppression of motions in the direction of gravity being an additional
constraint for the development of an instability. 
Hence, the neglect of this effect in
the present study justifies the more our conclusions about the stabilizing action of
rapid rotation. This could, however, imply that the marginal $P$, which 
discriminates between stability and instability, shifts to somewhat larger values.
In principle, compressibility, which is the cause for stratification, affects any MHD process
and may give rise to specific instabilities, e.g. the buoyancy (Parker) one, not existent
in an incompressible setup. Thus, deviation from the assumption of uniform density may add both a 
stabilizing and a destabilizing tendency. Although we are, guided by our estimate of the ratio of sound
speed to fluid velocity, quite sure that the stabilizing effect prevails, a quantitative certainty
can only be gained on the basis of a numerical study with an (at least) anelastic model.  

Apart from the magnetic field the vigorous flows occurring in the unstable situations may result in observable consequences: When extrapolating 
the maximum velocities found in the simulations to lower, more realistic viscosities, it seems likely that in some regions of the NS velocities
of $\ga 10^8$ cms$^{-1}$ can be reached. Supposed that the initial perturbation for the instability considered in this paper  is given by the 
(decaying) PNS convection, it is conceivable that the geometry of the latter is somehow ``transferred'' to the one of the Lorentz--force induced flow.
Moreover, as the rotation rates of the PNS and the newly born NS are almost identical,  it is allowed to speculate that similar to the PNS convection the demonstrated unstable decay of a strong magnetic field may be a source of gravitational waves.  Applying likewise the estimates of \cite{MJ97}, we conclude that for a duration from seconds to a few minutes a detectable gravitational 
wave signal can be expected rising at about 20\ldots 30 s after bounce.


Our study is based on the assumption, that newly born NSs inherit from their 
proto--NS phase surface magnetic fields  $\gtrsim10^{15}$G. This state can be reached either by flux conservation if the progenitor was already highly magnetized, by a PNS dynamo, or by other, still unknown, processes.

If the NSs are born with a strong dipolar field then
\begin{enumerate}
\item those whose rotation period is less than $\sim 6$ms and whose magnetic inclination angle $\alpha$ is smaller than $\sim45^\circ$ will retain their extremely large surface field strength and appear, after a rapid spin down, as magnetars;
\item those which rotate less rapidly, say with $P \ga 6$ms and/or for which is $\alpha \ga 45^\circ$ will lose almost all of their inborn magnetic energy and appear as radio pulsars.
\end{enumerate}
It turns out that rotation is likely to be the only stabilizing agent which allows of the existence of magnetars whereas the stable configurations found by \citet{BS05a} are less
suited to support such strong surface fields.

In the light  of the present study, the much smaller number of observed
magnetars in comparison with that of all the other observed realizations of NSs
seems to be a consequence of the fact that only a small fraction of all
new--born NSs are rotating as fast as or faster than $P\sim6$ms. Of course, this conclusion
is still premature, lacks broad observational support, does not consider other
selection effects, and has to be confirmed by the study of both more realistic
NS models and a denser mesh of initial $\alpha$ and $P$ values as well as by
studies of the evolutionary stages `supernova progenitor' -- `supernova' --
`proto--NS'. These considerations should definitely include the effects of magnetic fields and return a better idea about the distribution of magnetic field strengths, rotation periods and inclination angles at the beginning of a fully evolved NS's life.

Let us now relax the assumption that neutron stars have at the end of the PNS phase a field of $\ga 10^{15}$ G 
and adopt instead fields at that moment which are two to three orders of magnitude smaller. When additionally conceding that the
initial rotation period could be as large as, say, 60 ms, the ratio $q_P$ would be still at most $\sim 10^{-2}$.  As our results demonstrate,
such fields would be always stable irrespective of the magnetic inclination $\alpha$. 

\begin{acknowledgements}
This research was supported in part by the National Science Foundation under Grant No. PHY99-07949.
We are grateful to the organizers of the program ``The Supernova Gamma--Ray--Burst Connection'' of the Kavli Institute for Theoretical
Physics at the UCSB.
We thank F. Stefani for having made us aware of the Levitron\textsuperscript{TM} device.
\end{acknowledgements}
\bibliographystyle{aa}

\end{document}